\documentclass[conference]{IEEEtran}
\IEEEoverridecommandlockouts
\newcommand{\eat}[1]{}
\usepackage{cite}
\usepackage{subfigure}
\usepackage{amsmath,amssymb,amsfonts}
\usepackage{times}
\usepackage{algorithmic}
\usepackage{graphicx}
\usepackage{textcomp}
\usepackage{xcolor}
\usepackage{listings}
\usepackage{multirow}
\usepackage{multicol}
\usepackage{url}
\usepackage[pdftex]{hyperref}

\def\BibTeX{{\rm B\kern-.05em{\sc i\kern-.025em b}\kern-.08em
    T\kern-.1667em\lower.7ex\hbox{E}\kern-.125emX}}

\begin{document}

\title{Survive the Schema Changes: Integration of Unmanaged Data Using Deep Learning}

\author{\IEEEauthorblockN{Zijie Wang\IEEEauthorrefmark{1},
Lixi Zhou\IEEEauthorrefmark{1}, Amitabh Das\IEEEauthorrefmark{1},
Valay Dave\IEEEauthorrefmark{1}, Zhanpeng Jin\IEEEauthorrefmark{2}, and Jia Zou\IEEEauthorrefmark{1} }
\IEEEauthorblockA{\IEEEauthorrefmark{1}Arizona State University, \IEEEauthorrefmark{2}University at Buffalo, State University of New York \\
Email: \IEEEauthorrefmark{1}\{zijiewang, lixi.zhou, adas59, vddave, jia.zou\}@asu.edu,
\IEEEauthorrefmark{2}zjin@buffalo.edu}
}
\maketitle

\begin{abstract}
Data is the king in the age of AI. However data integration is often a laborious task that is hard to automate. Schema change is one significant obstacle to the automation of the end-to-end data integration process. Although there exist mechanisms such as query discovery and schema modification language to handle the problem, these approaches can only work with the assumption that the schema is maintained by a database. However, we observe diversified schema changes in heterogeneous data and open data, most of which has no schema defined. In this work, we propose to use deep learning to automatically deal with schema changes through a super cell representation and automatic injection of perturbations to the training data to make the model robust to schema changes. 
Our experimental results demonstrate that our proposed approach is effective for two real-world data integration scenarios: coronavirus data integration, and machine log integration. 

\end{abstract}


\lstnewenvironment{DSL}
  {\lstset{
        aboveskip=5pt,
        belowskip=5pt,
        mathescape=true,
        basicstyle=\ttfamily\small,
        morekeywords={Set, Vector, Map, HashMap, bool, select,
  multiselect, aggregate, join, partition, member, method, construct, true, false, if, else, CREATE TABLE, PARTITION BY,
  self, literal, vector, return, for push_back, function, enum, sort},
        literate={~} {$\sim$}{1},
        showstringspaces=false}\vspace{0pt}%
   \noindent\minipage{0.47\textwidth}}
   {\endminipage\vspace{0pt}}

\lstnewenvironment{SQL}
  {\lstset{
        aboveskip=5pt,
        belowskip=5pt,
        escapechar=!,
        mathescape=true,
        basicstyle=\linespread{0.94}\ttfamily\small,
        morekeywords={JOIN, FROM, WHERE, SELECT, AND},
        deletekeywords={VALUE, PRIOR},
        showstringspaces=false}
        \vspace{0pt}%
        \noindent\minipage{0.47\textwidth}}
  {\endminipage\vspace{0pt}}
\section{Introduction}
\label{sec:intro}
It was reported in 2018 that data scientists spent $80$-$90$\% efforts in the data integration process~\cite{abadi2020seattle, stonebraker2018data, thirumuruganathan2018data}. The schema change, which impacts applications and causes system downtimes, is always a main factor leading to the tremendous human resource overhead required for data integration. Schema changes are often caused by software evolution that is pervasive and persistent in agile development ~\cite{howard2011data}, or the diversity in data formats due to the lack of standards~\cite{campbell2016plea}. 

\begin{figure*}
\centering
\subfigure[A human-centered approach]{%
   \label{fig:human}
   \includegraphics[width=3.6in]{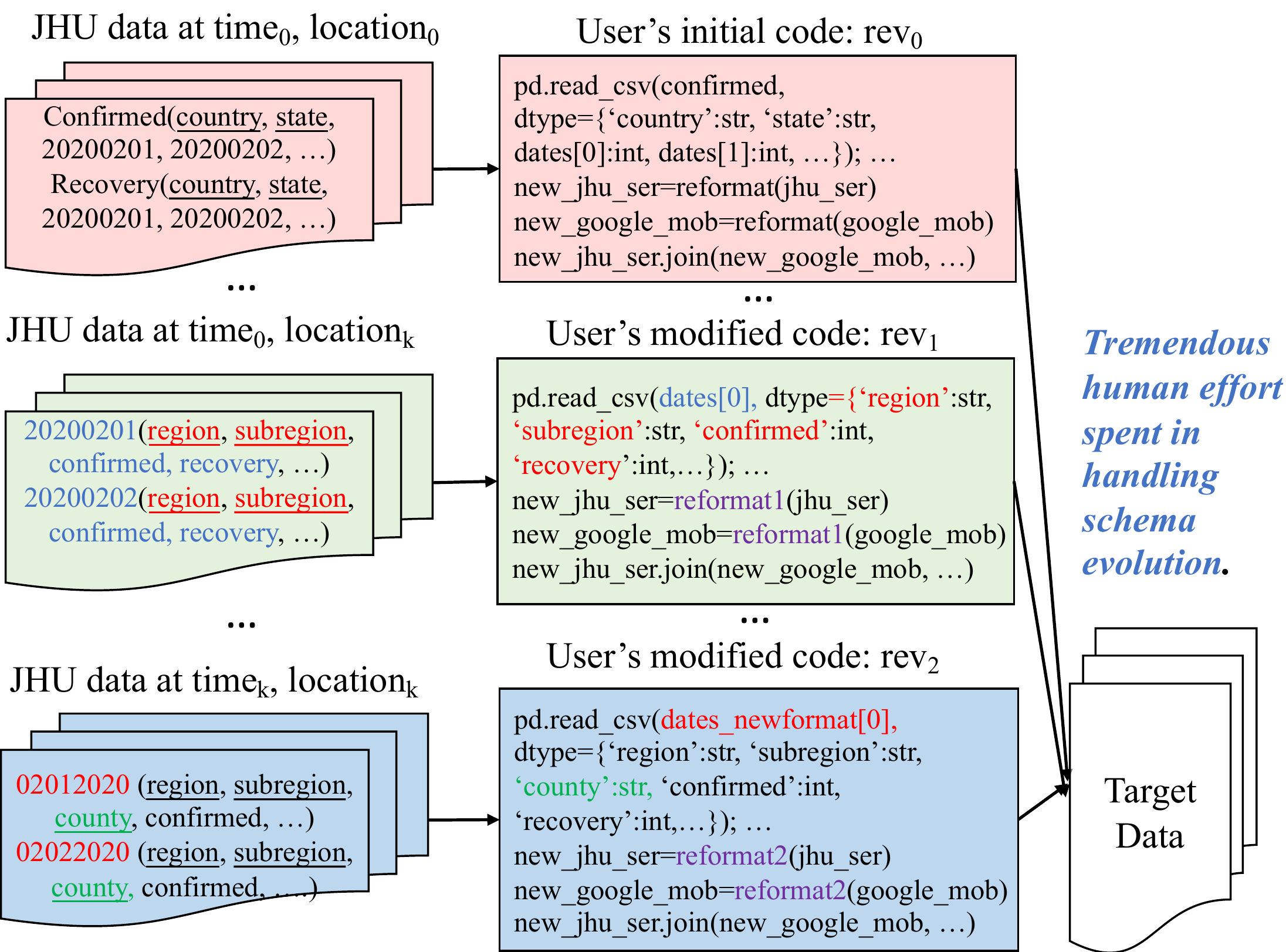}  
}%
\subfigure[A deep learning approach (our proposal)]{%
  \label{fig:automatic}
  \includegraphics[width=3.1in]{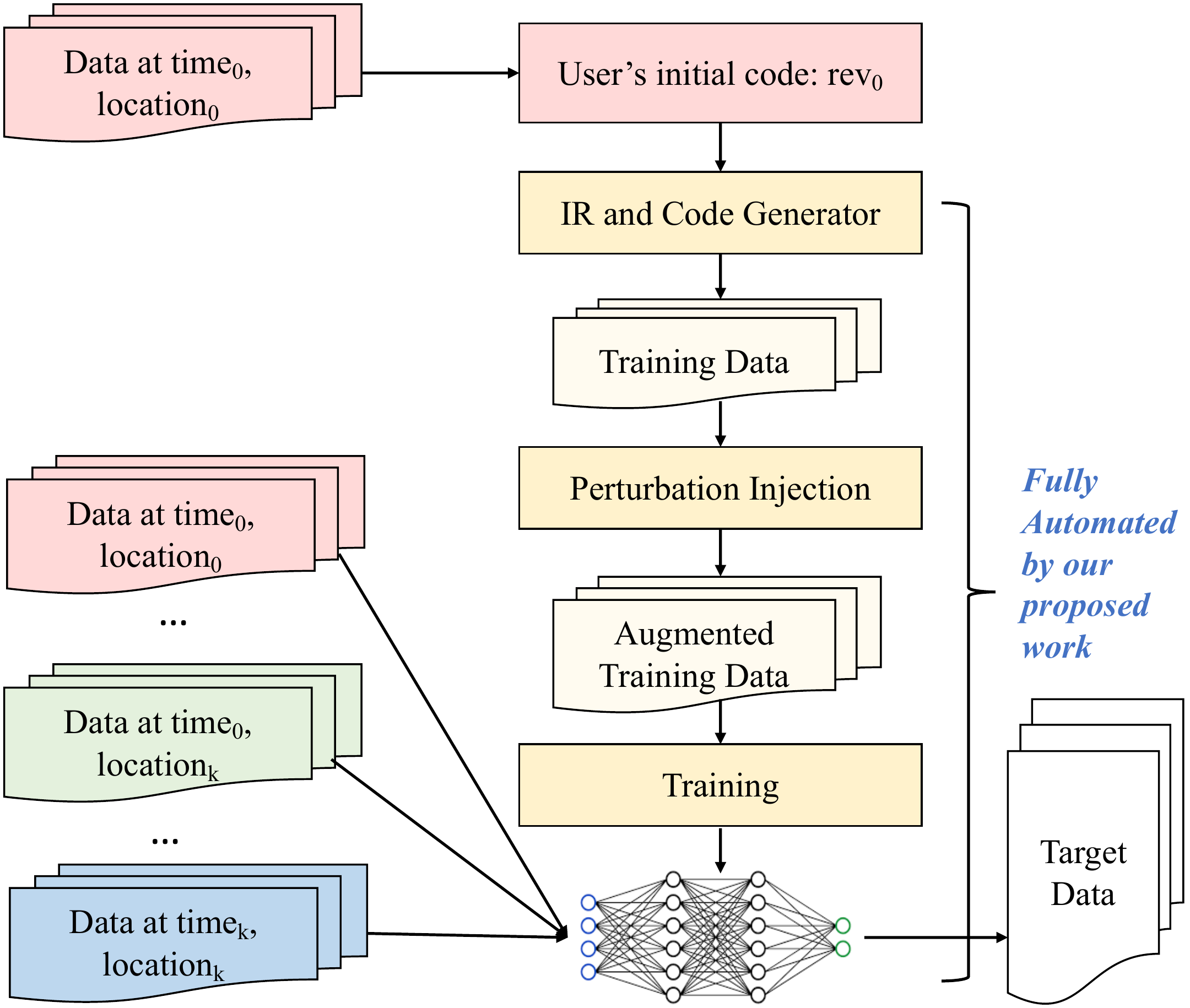}
}
\caption{\label{fig:motivation}
Motivation and Overview of Our Proposed Approach.
}
\end{figure*}

\noindent
\textbf{Example: coronavirus disease 2019 (COVID-19) data integration.} To predict the coronavirus outbreak, we integrate the coronavirus data repository at Johns Hopkins University (JHU) ~\cite{covid19-jhu} and the Google mobility data~\cite{google-mobility}. The JHU's data repository maintains the world coronavirus cases on a daily basis as a set of CSV files. However, we find the schema of the data has frequent changes. As illustrated in Fig.~\ref{fig:human}, the changes include attribute name changes (e.g., Longitude $\rightarrow$ Long\_), addition and removal of attributes (e.g., from six attributes initially to $15$ attributes), attribute type changes (e.g., date formats), key changes (e.g., from (\underline{country/region}, \underline{province/state}) to (\underline{combined\_key}), (\underline{FIPS}) and  (\underline{country\_region}, \underline{province\_state}, \underline{Admin2}). The data scientist's Python code for parsing these files easily breaks at each schema change, and requires manual efforts to debug and fix the issues. This becomes a ubiquitous pain for the users of this JHU data repository, and people even launched a project to periodically clean the JHU coronavirus data into a stable R-friendly format~\footnote{https://github.com/Lucas-Czarnecki/COVID-19-CLEANED-JHUCSSE}. However such cleaning is purely based on manual efforts and obviously not scalable.

\noindent
\textbf{Prior arts.} Schema evolution for data that was managed in relational databases, NoSQL databases, and multi-model databases are well-established research topics. The fundamental idea is to capture the semantic mappings between the old and the new schemas, so that the legacy queries can be transformed and/or legacy data can be migrated to work with the new schemas. There are two general approaches to capture the semantics mappings: (1) To search for the queries that can transform the old schema to the new schema~\cite{miller2000schema, hernandez2001clio, popa2002mapping, an2007semantic, fagin2009clio, shen2014discovering}. (2) To ask the database administrators (DBAs) or application developers to use a domain-specific language (DSL) to describe the schema transformation process~\cite{bernstein2007model, herrmann2017living, curino2013automating, moon2009prima, herrmann2015codel, hillenbrand2019migcast, moller2019query, klettkeevolution, moon2010scalable, scherzinger2013managing}.  

However, these approaches are not applicable to unmanaged data, including open data such as publicly available CSV, JSON, HTML, or text files that can be downloaded from a URL, and transient data that is directly collected from sensor devices or machines in real-time and discarded after being integrated. That's because the history of schema changes for these data is totally lost or become opaque to the users\eat{, unless we store such data back to a data management system, which could incur even more overhead than rewriting applications manually to handle system downtimes}. 
It is an urgent need to automatically handle schema changes for unmanaged data without interruptions to applications and any human interventions. Otherwise, with the rapid increase of the volume and diversity of unmanaged data in the era of Big Data and Internet of Things (IoT), it is unavoidable to waste a huge amount of time and human resources in manually handling the system downtimes incurred by schema changes.

\noindent
\textbf{A deep learning approach.} In this work, we argue for a new data integration pipeline that uses deep learning to avoid interruptions caused by the schema changes. In the past few years, deep learning (DL) has become the most popular direction in machine learning and artificial intelligence~\cite{lecun2015deep, schmidhuber2015deep}, and has transformed a lot of research areas, such as image recognition, computer vision, speech recognition, natural language processing, etc.. In recent years, DL has been applied to database systems and applications to facilitate parameter tuning~\cite{van2017automatic, sullivan2004using, li2019qtune, zhang2019end}, indexing~\cite{kraska2018case, ding2020alex}, partitioning~\cite{zou2020lachesis, DBLP:conf/sigmod/HilprechtBR20}, cardinality estimation and query optimization~\cite{kipf2018learned, krishnan2018learning}, and entity matching~\cite{mudgal2018deep, kasai2019low, konda2016magellan,thirumuruganathan2018reuse, zhao2019auto, ebraheem2017deeper}. While predictions based on deep learning cannot guarantee correctness, in the Big Data era, errors in data integration are usually tolerable as long as most of the data is correct, which is another motivation of our work. To the best of our knowledge, we are the first to apply deep learning to learn the process of join/union/aggregation-like operations with schema changes occurring in the data sources. However, it's not an easy task and the specific research questions include:

\noindent
(1) It is not straightforward to formulate a data integration task, which is usually represented as a combination of relational or dataflow operators such as \texttt{join}, \texttt{union}, \texttt{filter}, \texttt{map}, \texttt{flatmap}, \texttt{aggregate}, into a prediction task. What are effective representations for the features and labels?

\noindent
(2) How to design the training process to make the model robust to schema changes?

\noindent
(3) Different model architectures, for example, simple and compact sequence models such as Bi-LSTM and complex and large transformer such as GPT-2 and BERT, may strike different trade-offs among accuracy, latency, and resource consumption. What are the implications for model architecture selection in different deploying environments?

\noindent
(4) Annotating data to prepare for training data is always a major bottleneck in the end-to-end lifecycle of model deployment for production. Then how to automate training data preparation for the aforementioned prediction tasks?

\noindent
\textbf{Uninterruptible integration of fast-evolving data.} In this work, we first formulate a data integration problem as a deep learning model that predicts the position in the target dataset for each group of related data items in the source datasets. We propose to group related items in the same tuple or object that will always be processed together, and abstract each group into a super cell concept. We further propose to use source keys and attributes as features for describing the context of each cell, and use the target keys and attributes as labels to describe the target position where the super cell is mapped to. The features and labels can be represented as sentences or sentences with masks so that the representation can be applicable to state-of-art language models, including sequence models like Bi-LSTM and transformers like GPT-2 and BERT.

Then, to seamlessly handle various schema changes, inspired by adversarial attacks~\cite{ganin2016domain, kurakin2016adversarial}, which is a hot topic in DL, we see most types of schema changes as obfuscations injected to testing samples at inference time, which may confuse the model that is trained without noises. Therefore, just like adversarial training~\cite{shrivastava2017learning, ganin2016domain, kurakin2016adversarial}, we address the problem by adding specially designed noises to the training samples to make the model robust to schema changes. The techniques we employ to do so include replacing words by randomly changed words and synonyms that are sampled from Google Knowledge Graph. In addition, we propose to add an aggregation mode label to indicate how to handle super cells that are mapped to the same position, which can well handle the schema change of the type \textit{key expansion} in the example we give earlier.

Based on the above discussions, we propose a fully automated end-to-end process for uninterruptible integration of fast-evolving data sources, as illustrated in Fig.~\ref{fig:automatic}. \textit{Step 1}, the system will leverage our proposed Lachesis intermediate representation (IR)~\cite{zou2020lachesis} to automatically translate the user's initial data integration code into the executable code that automatically creates training data based on our proposed representation. \textit{Step 2}, obfuscations will be automatically injected to the training data to make the model robust to various schema changes. \textit{Step 3}, different model architectures will be chosen to train the predictive model that will perform the data integration task depending on the model deploying environment. Due to the space limitation, this paper will focus on Step 2 and 3, while we will also discuss Step 1 as well as other techniques for reducing training data preparation overhead such as crowdsourcing and model reuse.

The contributions of this work include:

\noindent
(1) As to our best knowledge, we are the first to systematically investigate the application of deep learning and adversarial training techniques to automatically handle schema changes occurring in the data sources.

\noindent
(2) We propose an effective formulation of the data integration problem into a prediction task as well as flexible feature representation based on our super cell concept (Sec.~\ref{sec:formulation}). We also discuss how to alleviate the human costs involved in preparing training data for the representation (Sec.~\ref{sec:automation}).

\noindent
(3) We represent the common schema changes as various types of obfuscations, which can be automatically injected to the training data to make the model training process robust to these types of schema changes. (Sec.~\ref{sec:perturbations})  

\noindent
(4) We compare and evaluate various trade-offs made by different model architectures, including both simple sequence model and complex transformer model for two different data integration tasks involving semi-structured and non-structured data respectively. (Sec.~\ref{sec:training} and Sec.~\ref{sec:experiments})  

\section{Problem Analysis and Formulation}
\label{sec:formulation}
\subsection{Assumptions}
We assume that in a typical data integration scenario that converts a set of source datasets into one target dataset, each of the source datasets may have heterogeneous formats such as CSV, JSON, text, etc., which may not be managed by any relational or NoSQL data stores; However, the target dataset must be tabular, so that each cell in the target dataset can be uniquely identified by its tuple identifier and attribute name. 

\subsection{Representation Analysis and Comparison}
Given a set of objects, where each object may represent a row in a CSV file, a JSON object, a time series record, or a file, there are a few candidate representations for formulating the predictive task, including dataset-level representation, object-level representation, attribute-level representation, cell-level representation, and our proposed super cell based representation. The coarser-grained of the representations, the fewer times of inferences are required, and the more efficient of the prediction. However, a coarser-grained representation also indicates that the prediction task is more complex and harder to train a model with acceptable accuracy, because a training sample will be larger and more complex than other finer-grained representations, and the mapping relationship to learn will naturally become more complicated.

\begin{figure} [h]
\centering
   \includegraphics[width=3.5in]{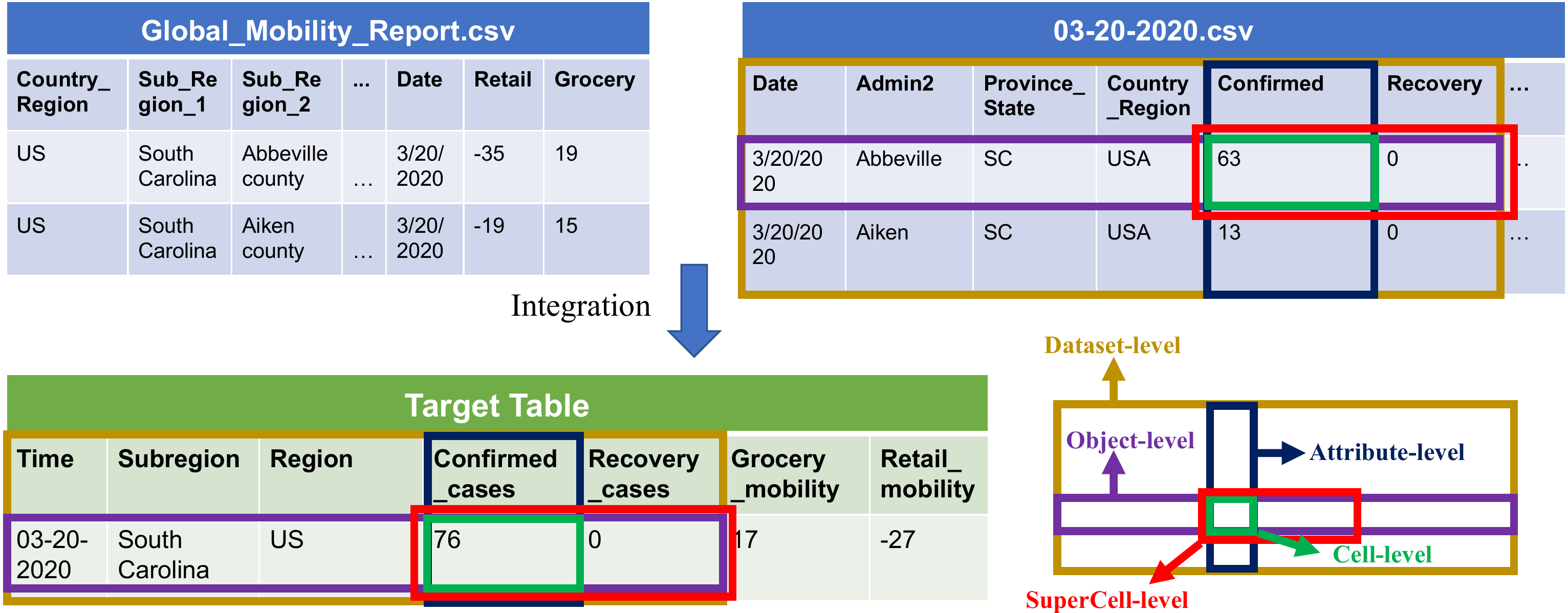}
\caption{\label{fig:representations}Various representations at different granularities}
\end{figure}

Various levels of representations for the motivating example are illustrated in Fig.~\ref{fig:representations}. We find it almost impossible to collect sufficient training data represented at the dataset-level. The object-level representation groups all attributes and is not expressive enough in describing the different transformations applied to each attribute. Similarly, the attribute-level representation assembles all values in the same attribute and is not efficient in expressing logics like filtering or aggregation. The cell-level representation, referring to a value of a single attribute in one tuple, is at the finest granularity, which simplifies the learning process. However, it may incur too many inferences and waste computational resources, particularly if there exist multiple cells in one object that will always be mapped/transformed together.

Therefore we propose and argue for a super cell representation. A super cell is a group of cells in an object that will always be mapped to the target table together in a similar way, such as the values of the \texttt{confirmed} attribute and the \texttt{recovery} attribute, as shown in Fig.~\ref{fig:representations}. While the flexible granularity of a super cell is between the object-level and cell-level, it can well balance the expressiveness and the performance regarding the training and testing process. 

For each super cell, we represent its features as a sentence that concatenates the keys, as well as the value and attribute of each cell in the super cell. 
Particularly the keys in the context features should include both of the join key if applicable and the tuple/object identifier of the local table. We find that for $1$-$1$ join, the join key usually also serves as a local tuple/object identifier; for $1$-$N$ and $N$-$M$ joins, the join key is not necessarily the local key; and for non-join operations (e.g., union, filter, aggregate), no join key is required. We assume the target dataset is always tabular and propose a label representation that includes the tuple identifier (i.e., key of the target table) of the super cell, the target attribute name of each cell, and an aggregation mode to specify how values that are mapped to the same position should be aggregated into one value, e.g., add, avg, max, min, count, replace\_old, discard\_new, etc..

\noindent
\textbf{Super cell formulation.} Suppose there are $m$ source datasets, represented as $D = \{d_i\} (0\leq i<m)$, each source dataset is modeled as a set of $n_i$ super cells, denoted as $d_i = \{s_{ij}\} (0\leq i<m, 0\leq j<n_i)$. We further describe each super cell $s_{ij} \in d_i$ as a triplet that consists of three vectors for the keys shared by each cell in the super cell, attribute names of each cell, and values of each cell, respectively, represented as  $s_{ij} = (\vec{key}_{ij}, \vec{attribute}_{ij}, \vec{value}_{ij})$. We can further define a super set to describe the current state of the entire data repository: $S=\{s_{ij}\in d_i | \forall d_i \in D\}$. 

\begin{figure} [h]
\centering
   \includegraphics[width=3.5in]{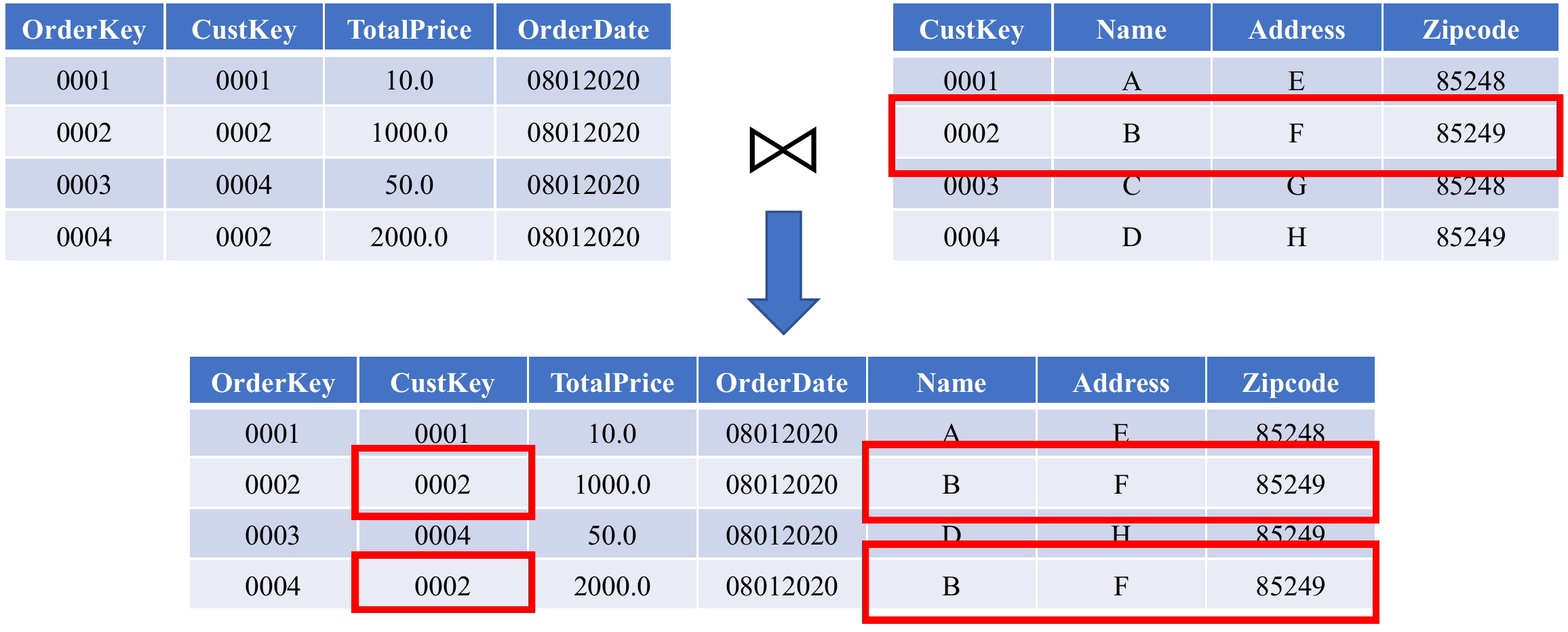}
\caption{\label{fig:1toN}Example of 1-N join}
\end{figure}

A super cell may be mapped to zero, one, or more than one positions in the target dataset, depending on the operations involved in the data integration task. For example, as illustrated in Fig.~\ref{fig:1toN}, in a $1$-$N$ join operation, a super cell in the table at the left-hand side may be mapped to many positions in the target table. Thus we can represent the target positions where the super cell is mapped to as a list of triples:  $f(s_{ij})=\{(\vec{key^T}_{ij}, \vec{attribute^T}_{ij}, agg\_mode)\}$, where each triple refers to one position that is indexed by the target keys $\vec{key^T}$ shared by all cells in the super cell, as well as attribute name of each cell in the super cell, denoted as $\vec{attribute^T}$.

\eat{
\begin{figure} [h]
\centering
   \includegraphics[width=3.2in]{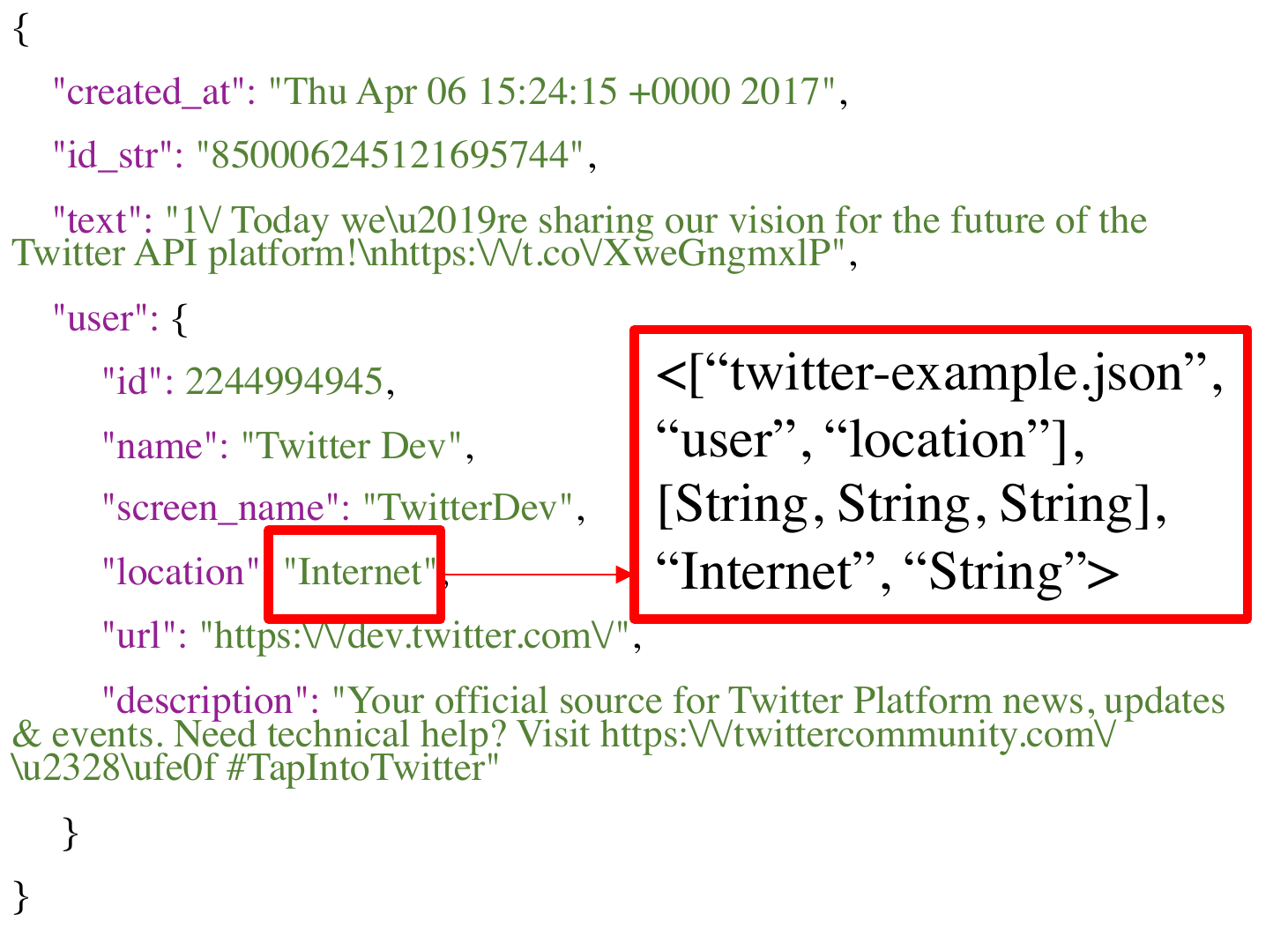}
\caption{\label{fig:json-representation}
Key-value representation for JSON object}
\end{figure}

\begin{figure} [h]
\centering
   \includegraphics[width=2.5in]{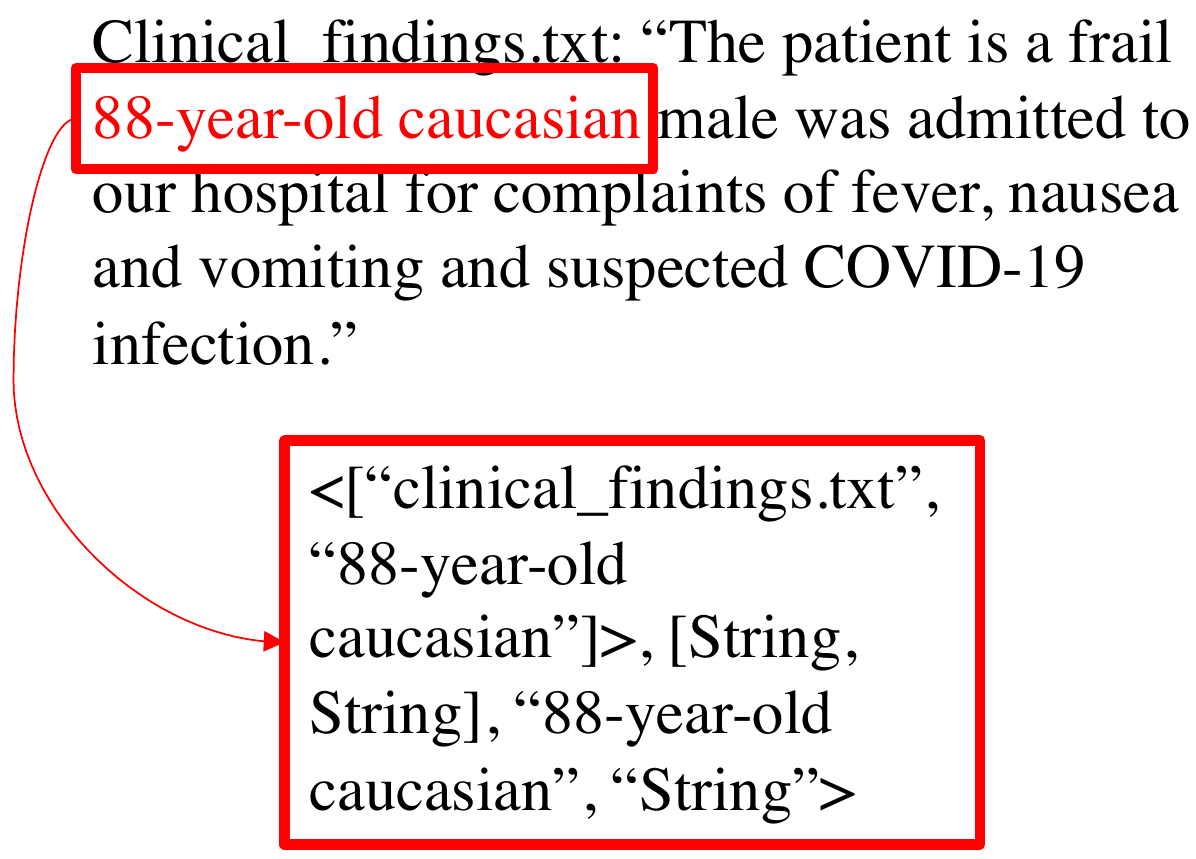}
\caption{\label{fig:text-representation}
Key-value representation for textual file}
\end{figure}
}

\subsection{Problem Definition}
Given a fast evolving data repository $S=\{s_{ij}\in d_i | \forall d_i \in D\}$, for a data integration request, the user input should specify the schema of the expected target table. The schema includes a list of $p$ attributes denoted as $A = \{a_{k}\} (0\leq k<p)$, where $a_k$ represents the $k$-th attribute in the target table, as well as a list $q$ attributes that serve as minimum tuple identifier (i.e. key) denoted as $A^{key} = \{a_{l}\} (0\leq l<q)$. We further denote a set of all possible key values in the target table as $R=\{r_l | r_l \in a_l\} (0\leq l<q)$. Then we need find a model $f^{S\rightarrow (A\cup\{NULL\})\times (R\cup\{NULL\})}$ that predicts a set of target positions denoted as $\{(\vec{key^T}_{ij}, \vec{attribute^T}_{ij}, agg\_mode)\}$ for each super cell $s_{ij} \in S$, where each element of the key $\forall x \in [0, q), \vec{key^T}_{ij}[x] \in R$, and each element of the attribute vector $\forall y \in [0, |s_{ij}|),\vec{attribute^T}_{ij}[y] \in A$, where $|s_{ij}|$ denotes the number of cells in the super cell $s_{ij}$. If a super cell doesn't belong to the target table and should be discarded, we define that $r_l=NULL$ and $a_k=NULL$ in this case.

\section{Training Data Augmentations for Schema Changes}
\label{sec:perturbations}
We identify five basic types of data schema changes, which cover all of the relational or NoSQL schema changing patterns ~\cite{bernstein2007model, herrmann2017living, curino2013automating, moon2009prima, herrmann2015codel, hillenbrand2019migcast, moller2019query, klettkeevolution, moon2010scalable, scherzinger2013managing} as well as schema changes that we have discovered from open data. We first discuss the impact of each type and then propose our approaches to handle these changes by adding perturbations to the training process.

\noindent
(1) Domain pivoting. For example, originally the dataset was stored as three CSV files, describing the daily confirmed coronavirus cases, daily recovery cases, and daily death cases; later the schema changed to a new set of CSV files so that each file described all information (confirmed, death, recovery cases) on that specific date. We observed such changes prevalent in the coronavirus data~\cite{covid19-jhu,google-mobility,covid19-harvard} and the weather database hosted by National Oceanic and Atmospheric Administration (NOAA)~\cite{noaa}. Such changes will easily break a conventional Python-based data integration pipeline.

\noindent
(2) Key expansion. For example, the key of the dataset is lowered down from the state level (\underline{country/region}, \underline{province/state}) to the county level (\underline{combined\_key}), (\underline{FIPS}) and  (\underline{country\_region}, \underline{province\_state}, \underline{Admin2}), which means a tuple in the original table (that describes statistics for a state) is broken down into multiple tuples with each describes the statistics for a county. Such changes cannot be easily handled by conventional data integration methodologies.

\noindent
(3) Attribute name and ordering change. For example, in the first CSV file added to the JHU COVID-19 daily report data repository on Jan 22nd, 2020, the third column name is ``Last Update". But in the CSV file added to the same repository on Sept 24th, 2020, the same column is moved to the fifth position, and the name is slightly changed to ``Last\_Update". Such changes may interrupt a conventional program that joins two datasets on the ``Last Update" column.

\noindent
(4) Value type/format change. For example, in the daily COVID-19 file created on Jan 22nd, the "Last Update" has values in the format of ``1/22/2020 17:00". However, in the file on Sept 24th, the value format has changed to ``2020-09-25 04:23:21". A conventional exact join operation using "Last Update" as the join key cannot handle such value format change, unless the programmer chooses to convert it into a similarity join, which is more complicated and much slower than an exact join and will result in significantly higher development costs~\cite{zhu2017auto,xiao2008ed}.

\noindent
(5) Addition or removal of non-key attributes. For example, the JHU COVID-19 global daily report has changed from six attributes initially to $14$ attributes after a few months. This change may not make much influence, if the affected attributes are not used by users' data integration workloads. On the contrary, if a required column is totally removed, there is no way to handle such a situation without interruption, even if using a deep learning approach, so we mainly focus on the first four types of schema changes.

\noindent
\textbf{Perturbation based on schema changes.} First, our super cell based representation will not be affected by dimension pivoting, attribute ordering change, and addition or removal of irrelevant attributes, because the context for any cell remains the same despite of these schema changes. Therefore a model trained with our proposed representations is robust to these types of schema changes. 

Second, schema changes such as renaming of an attribute and reformatting of cell values, are similar to adversarial attacks, which confuse the pre-trained models by adding noises to the expected testing samples.  Adding perturbations to training data is an effective way of training robust models against adversarial attacks~\cite{ganin2016domain}. Inspired by this analogy, we add specially designed perturbations to training data to handle these parts of schema changes. If we see each super cell representation as a sequence of words (i.e., a sentence), the training data is a corpus of sentences. Then we can augment the training data by adding new sentences (i.e., perturbations), which are changed from existing sentences by randomly replacing a word using synonyms extracted from Google Knowledge Graph~\cite{covid19-jhu, google-mobility, covid19-harvard}, or randomly modified words by removing one or more letters. Then we train a character-based embedding on this augmented corpus using fastText~\cite{fastText}, which  maps related words to vectors that are close to each other so that the model can recognize the similarity of such words. We observe through experiments that character-based embedding can achieve better accuracy and reliability than word-based embedding and can smoothly handle out-of-vocabulary words. It also shows that our locally trained embedding significantly outperforms pre-trained embeddings with Google News or Wikipedia.

Third, to make the deep learning model robust to key expansion, as mentioned, we add a new label to the representation called ``aggregation mode". Each value of the label represents an aggregation operator such as \texttt{sum}, \texttt{avg}, \texttt{min}, \texttt{max}, which will be applied to the cells that are mapped to the target position; \texttt{replace}, which means the cell will replace the old cell that exists in the same position of the target table; or \texttt{discard}, which means the new cell will be discarded if an older cell has been mapped to the same target position. 
\section{Model Training and Inference}
\label{sec:training}

\subsection{Model Architectures}
In recent several years, Natural Language Processing (NLP) has experienced several major advancements including the bi-directional mechanism, attention mechanism, transformer mechanism, and so on. Existing works show that the final hidden state in Bi-LSTM networks cannot capture all important information in a long sentence. Therefore, the attention mechanism was introduced to address the problem by preserving information from all hidden states from encoder cells and aligning them with the current target output. Later such idea was integrated into the transformer architectures, so that encoders had self-attention layers, while decoders had encoder-decoder attention layers. Most recently, to make the transformer architecture more flexible to applications other than language translation, GPT-2 that only uses the decoders' part and BERT that only uses encoders' part are invented and achieve great success in a broad class of NLP problems.
Our assumption is that on one hand, more complicated models like GPT-2 and BERT may naturally achieve better accuracy than a simpler model like Bi-LSTM; but on the other hand, these complex models may require significantly higher storage and computational resources, as well as more training data. It is important to know the trade-offs among accuracy, latency, and resource consumption, made by different model architectures.
We mainly consider two types of language model architectures: (1) simple and compact sequence models based on customized local character-based embedding and Bi-LSTM; and (2) complex and large pre-trained transformer models, such as GPT-2~\cite{radford2019language} and BERT~\cite{devlin2018bert}. 

\subsubsection{Sequence Model (Bi-LSTM)}
Our Bi-LSTM model architecture, includes an embedding layer that has $150$ neurons; a Bi-LSTM layer that consists of $512$ neurons; and a fully-connected layer that has $256$ neurons. 

\subsubsection{Transformer Model}
 Moreover, we also consider transformer models based on GPT-2~\cite{radford2019language} and BERT~\cite{devlin2018bert}. We use a pre-trained GPT-2 small model or a pre-trained BERT base model as the backend, which connects to a frontend classifier composed of four convolutional layers and a fully connected layer. During the training process, the parameters of the GPT-2 small model and the BERT base model are freezed, and only the parameters of the frontend will be updated.
 
The pre-trained GPT-2 small has $117$ millions of parameters, including $12$ layers of transformers, each with 12 independent attention mechanisms, called “heads”, and an embedding size of $768$ dimensions. The hidden vector output from the GPT-2 small model is reshaped to add a channel dimension and then passed to four convolutional layers, including two max-pooled 2D convolution layer and two average-pooled 2D convolution layer respectively, the output is applied with a hadamard product, and then sent to a fully connected layer. The BERT base model has $110$ millions of parameters, with $12$ transformer blocks, and each has $768$ hidden neurons and $12$ self-attention heads. It uses the same architecture of the frontend classifier with the GPT-2 small model. 
Although GPT-2 and BERT are both based on the transformer model, they use different units of the transformer. GPT-2 is built using transformer decoder blocks constructed by the masked self-attention layers, while the BERT utilizes transformer encoder blocks with self-attention layers. 

Although transformer models usually achieve better accuracy than sequence models through its attention mechanism, they also require significantly more storage space. For example, GPT-2 small, which is the smallest variant of GPT-2 model requires more than $500$ megabytes of storage space; The BERT base model~\footnote{https://tfhub.dev/google/bert\_uncased\_L-12\_H-768\_A-12/1} that we use takes $450$ megabytes of storage space.
In contrast, the Bi-LSTM model is smaller than $1$ megabyte.

\subsection{Assembling of Inference Results}
For each super cell, the model will predict a set of target positions in the form of $\{(\vec{key^T}_{ij}, \vec{attribute^T}_{ij}, agg\_mode)\}$, as we mentioned in Sec.~\ref{sec:formulation}. Then based on each super cell and its predicted positions, a general data assembler will put each value to the right places in the target table. Based on the configuration, the assembler can work in either local mode by buffering and writing one file to store the target dataset in local or dispatch the assembled tuples to users' registered deep learning workers (i.e., target data is consumed by a deep learning application) once an in-memory buffer is full. In the latter case, in each deep learning worker's side, a client is responsible for receiving and assembling tuples into the target dataset. During the dispatching process, the output table will be partitioned in a way to guarantee load balance and ensure the independent identical distribution (i.e., IID) to avoid introducing bias. 
\section{Automation of Training Data Preparation}
\label{sec:automation}
\subsection{Code Generation Based on Intermediate Representation}
\label{sec:codegen}
An important objective of this work is to free human experts from all dirty works of wrangling with schema changes. Therefore it's critical to reduce the human efforts required in training data preparation, such as parsing and annotating data. We propose to automate the training data creation by utilizing conventional Python code developed for integrating an initial set of data sources. The users' Python codes specify how to transform the data sources (usually with heterogeneous formats) to a target table (usually in tabular format), which is exactly the information needed for creating the training data. This gives us an opportunity to translate users' data integration code to training data preparation code.

For relational data, the integration logic can be fully expressed in SQL, which maps to relational algebra. Then it is easy to generate code for training data creation process based on the relational algebra. First, all key and join key information are well maintained and can be directly retrieved. Second, it is easy to identify which attributes of a table will always be processed similarly by analyzing the relational algebra expression, so that the values of these attributes in the same tuple can be grouped into a super cell. \eat{For example, by analyzing a query such as \textit{SELECT COVID-19.Name, Customer.Address, Customer.ZipCode, Order.TotalPrice, Order.Date FROM Customer, Order WHERE Customer.orderID = Order.ID AND Order.TotalPrice $>$ 5000;}, we will know that \textit{Customer.Name}, \textit{Customer.Address}, and \textit{Customer.ZipCode} should be grouped into one super cell, and \textit{Order.TotalPrice}, \textit{Order.Date} should be grouped into another super cell. Third, it is possible to rewrite each relational operator, so that the input and output of the relational operators are all based on the super cell representation. For example, the input of the select operator of the above example is in form of \textit{((Order.ID), (Order.TotalPrice, Order.orderDate), (3000, 2020-10-05))}, and the output of the select operator is in form of $\langle$\textit{((Order.ID), (Order.TotalPrice, Order.orderDate), (3000, 2020-10-05)), false}$\rangle$, so that this super cell will be filtered out and serve as the input to the join operator. Then the join outputs are a set of pairs of matching super cells, such as $\langle$\textit{(("O000001"), (Order.TotalPrice, Order.orderDate), (6000, 2020-10-05))}, \textit{(("C1010101", "O000001"), (Customer.Name, Customer.Address, Customer.ZipCode), ("Steves", "9210 Main Street", 85248))}$\rangle$. This output can be easily transformed into training data.} 
For example, by analyzing a query coded up for a data integration task such as 

\begin{SQL}
SELECT COVID-19.Date, COVID-19.State,
COVID-19.Country, COVID-19.Confirmed, 
COVID-19.Recovered, Mobility.Workplace, 
Mobility.Recreation, Mobility.Grocery 
FROM COVID-19, Mobility 
WHERE COVID-19.Date = Mobility.Time 
AND COVID-19.Country = Mobility.Region 
AND COVID-19.State = Mobility.SubRegion;
\end{SQL} 

\noindent
, we will know that \textit{COVID-19.Confirmed} and \textit{COVID-19.Recovered} should be grouped into one super cell; and \textit{Mobility.Workplace}, \textit{Mobility.Recreation}, and \textit{Mobility.Grocery} should be grouped into another super cell. Third, it is possible to rewrite each relational operator, so that the input and output of the relational operators are all based on the super cell representation. For example, the input of the \textit{Join} operator in the above example has the form of \textit{((2020-10-06, AZ, US), (COVID-19.Confirmed, COVID-19.Recovered), ($3103$, $2214$))}. Then the join outputs are a set of pairs of super cells that match the join predicate, such as $\langle$\textit{((2020-10-06, AZ, US), (COVID-19.Confirmed, COVID-19.Recovered), ($3103$, $2214$))}, \textit{((2020-10-06, AZ, US), (Mobility.Workplace, Mobility.Recreation, Mobility.Grocery), ($21$, $5$, $17$))}$\rangle$. This output can be easily transformed into a base set of training data, into which the perturbations will be injected.

However, because the integration code of open data, is usually written in an object-oriented language such as Python, Java, C++, the code after compilation is opaque to the system, and it is hard to modify the code directly. One solution is to map the integration code to an intermediate representation (IR), such as Weld IR~\cite{palkar2017weld} that is integrated with libraries like numpy and SparkSQL; and our proposed Lachesis IR~\cite{zou2020lachesis}. Such IR is usually a directed acyclic graph (DAG), and can be reasoned by the system. In this DAG, each node is an atomic computation, and 
each edge represents a data flow or a control flow from the source node to the destination node. The atomic computations useful to data integration workloads usually can be composed by three categories of operators:

\noindent
(1) \textit{Lambda abstraction functions} such as a function that returns a literal (a constant numerical value or string), a member attribute or a member function from an object; unary functions such as \texttt{exp}, \texttt{log}, \texttt{sqrt}, \texttt{sin}, \texttt{cos}, \texttt{tan}, etc.\eat{; or opaque unary functions if the programmer prefers not to expose the logic}. 

\noindent
(2) \textit{Higher-order lambda composition functions} such as binary operators: \texttt{\&\&}, \texttt{||}, \texttt{\&}, \texttt{|}, \texttt{<},\texttt{>}, \texttt{==}, \texttt{+}, \texttt{-}, \texttt{*}, \texttt{/}; conditional operator like \texttt{condition? on\_true:on\_false}; etc..

\noindent
(3) \textit{Set-based operators} such as \texttt{scan} and \texttt{write} that reads/writes a set of objects from/to the storage; \eat{\texttt{partition} that shuffles a set of objects across a cluster of nodes; \texttt{replication} that replicates a set of objects to all cluster nodes;\texttt{apply} that applies a lambda calculus expression (i.e. composed of lambda abstractions and higher order composition functions) to a set of objects (like \textit{map});} \textit{map}, \texttt{join}, \texttt{aggregate}, \texttt{flatten}, \texttt{filter}, etc..

We propose to modify existing intermediate representations, so that a super cell based processor can be derived from each atomic computation. We assume that each source dataset can be represented as Pandas dataframes. Then by traversing the IR graph, the system can understand the keys and the super cell mapping relationship. The super cell based processor of each of atomic computations transforms each super cell representation accordingly. For example, \texttt{map} operator that transforms a date cell ``2020-10-06" to ``Oct 6, 2020" as an example, the processor takes a super cell \textit{\{``keys": [``2020-10-06", ``AZ", ``US"], ``attributes": [``Date"], ``cells":[``2020-10-06"]\}} as input, and outputs \textit{\{``keys": [``2020-10-06", ``AZ", ``US"], ``attributes": [``Date"], ``cells":[``Oct 6, 2020"]\}}  so that the contextual relationship between ``Oct 6, 2020" and its source key and attribute name is preserved. The \texttt{write}'s processor transforms each super cell into a $\langle{feature, label}\rangle$ representation, such as \textit{\{``source super cell":\{``keys": [``2020-10-06", ``AZ", ``US"], ``attributes": [``Date"], ``cells":[``Oct 6, 2020"]\}, ``target position": \{``keys": [``Oct 6, 2020", ``Arizona", ``United States"], ``attributes":[``datetime"]\}\}}.  In this way, we can obtain training data automatically.

\eat{
We propose to modify existing intermediate representations, so that an super cell based processor can be derived from each atomic computation. We assume that each input object can be represented as a set of cells (e.g. a set of attributes) with the cell(s) that serve as the object identifier specified, while a cell can be as complex as a dictionary or list, for example, Pandas dataframes, Spark dataframes and so on. Then by traversing the IR graph, the system can understand the super key relationship as well as the super cell mapping relationship. The super cell based processor of each of atomic computations transforms each super cell representations to a new super cell representations, using a set-oriented computation that is equivalent to the original atomic computation. For example, \texttt{map} operator that transform a date cell "2020-10-06" to "Oct 6, 2020" as an example, the processor takes input as \textit{((2020-10-06, AZ, US), (Date), ("2020-10-06"))}, and outputs \textit{((2020-10-06, "AZ", "US"), ("Date"), ("Oct 6, 2020"))} so that the mapping relationship between "Oct 6, 2020" and corresponding key and attribute name is reserved. The \texttt{write}'s processor transforms each target cell into a $\langle{feature, label}\rangle$ representation, such as \textit{$\langle$((2020-10-06, AZ, US), ("Date"), ("Oct 6, 2020")), (("Oct 6, 2020", "Arizona", "United States"), ("Time"))$\rangle$}.  This way, we can obtain training data automatically.}

However, the limitation of above approach is that it may not work if the input object is totally nested and opaque and cannot be represented as a set of cells like Pandas dataframes or Spark dataframes. For example, a corpus of totally unstructured text files, unavoidably requires human pre-processing efforts. Thereby, we design following approaches to further alleviate the problem: model reusing and crowdsourcing.

\subsection{Crowdsourcing} 
According to Sec.~\ref{sec:codegen}, if we are able to convert an unstructured dataset, e.g., a set of opaque and nested objects or a set of unstructured texts, into a Pandas dataframe or similar structures, the code generation approach maybe applicable to automate the training data creation process. However, it is non-trivial to identify the parsing logic, perform such conversion and identify the keys. All these tasks are hard to automate. We consider crowdsourcing as a potential approach to alleviate the burden from the data scientists or domain experts for these tasks. However, based on our experiments of crowdsourcing $160$ key identification tasks to $8$ graduate students, and $164$ undergraduate students from an introductory database course, requesting to identify all keys. We find that the accuracy is merely $65.7\%$. First, some of the datasets, particularly these scientific datasets, require domain-specific knowledge to tell the tuple identifier, because these attribute names are acronyms or terms that are not understandable to most people who are not in the domain, and usually datasets are not shipped with detailed explanations for each attribute. Second, for large datasets, it is impossible for a person who are not familiar with the datasets to tell the keys. Third, it is not easy to find a lot of people who has database knowledge. Other tasks such as identifying super cells and parsing unstructured datasets are even more challenging for crowdsourcing platforms due to the expert knowledge required in nature.

\subsection{Model Reuse}
Another approach to reduce human efforts involved in preparing training data is to reuse models for similar data integration tasks. For this purpose, we design and develop a system, called as ModelHub, which searches for reusable models for a new data integration task by comparing the attributes of the target dataset (that is created by the programmer's initial data integration code) with the target dataset of each existing data integration models. We leverage locality sensitive hashing (LSH) based on MinWise hash~\cite{zhu2016lsh, datar2004locality} for text-based data and LSH based on JS-divergence~\cite{chen2019locality, mao2017s2jsd} for numerical data to accelerate the attribute-matching process. Another benefit of utilizing the LSH is that, in the ModelHub platform, each model only needs to be uploaded with LSH signatures of the target dataset's attributes, while the target dataset does not need to be submitted, which saves the storage overhead and also addresses privacy concerns.

\section{Evaluation}
\label{sec:experiments}
We mainly answer following questions in this section:

\noindent
(1) How effective is our proposed deep learning representation for different data integration tasks?

\noindent
(2) How effective are the perturbations added to the training data for handling various types of schema changes?

\noindent
(3) How will different super cell granularities affect the accuracy, and the overheads for the training, testing, and assembling process?

\noindent
(4) How will different model architectures (complex and large models vs. simple and compact models) affect the accuracy and latency for different types of data integration tasks?

\noindent
(5) How will our approach of handling schema changes improve productivity and alleviate programmers' efforts?

\subsection{Environment Setup}
Based on the proposed training data representation and training data perturbation methodology, we have created training data to train Bi-LSTM model, GPT-2 small model, and BERT base model for two scenarios: coronavirus data integration and heterogeneous machine data integration. The first scenario mainly involves tabular source datasets in CSV formats with aforementioned schema changes. However, the source datasets for the second scenario are mainly unstructured text data, in which most of the similar terms in different platforms are expressed very differently (e.g., CPU user time is logged as ``CPU usage: $14.90\%$ user" in MacOS, ``\%Cpu(s): $14.9$ us" in Ubuntu, and ``$400$\%cpu  $86$\%user" in Android).

\noindent
\textbf{Model Architectures.} We compare three neural networks: Bi-LSTM, GPT-2 small with a CNN frontend classifier, and BERT base with the same CNN frontend classifier. The model architectures are described in Sec.~\ref{sec:training}. 

\begin{figure}[ht]
\centering\subfigure[Bi-LSTM model]{%
   \includegraphics[width=1.67in]{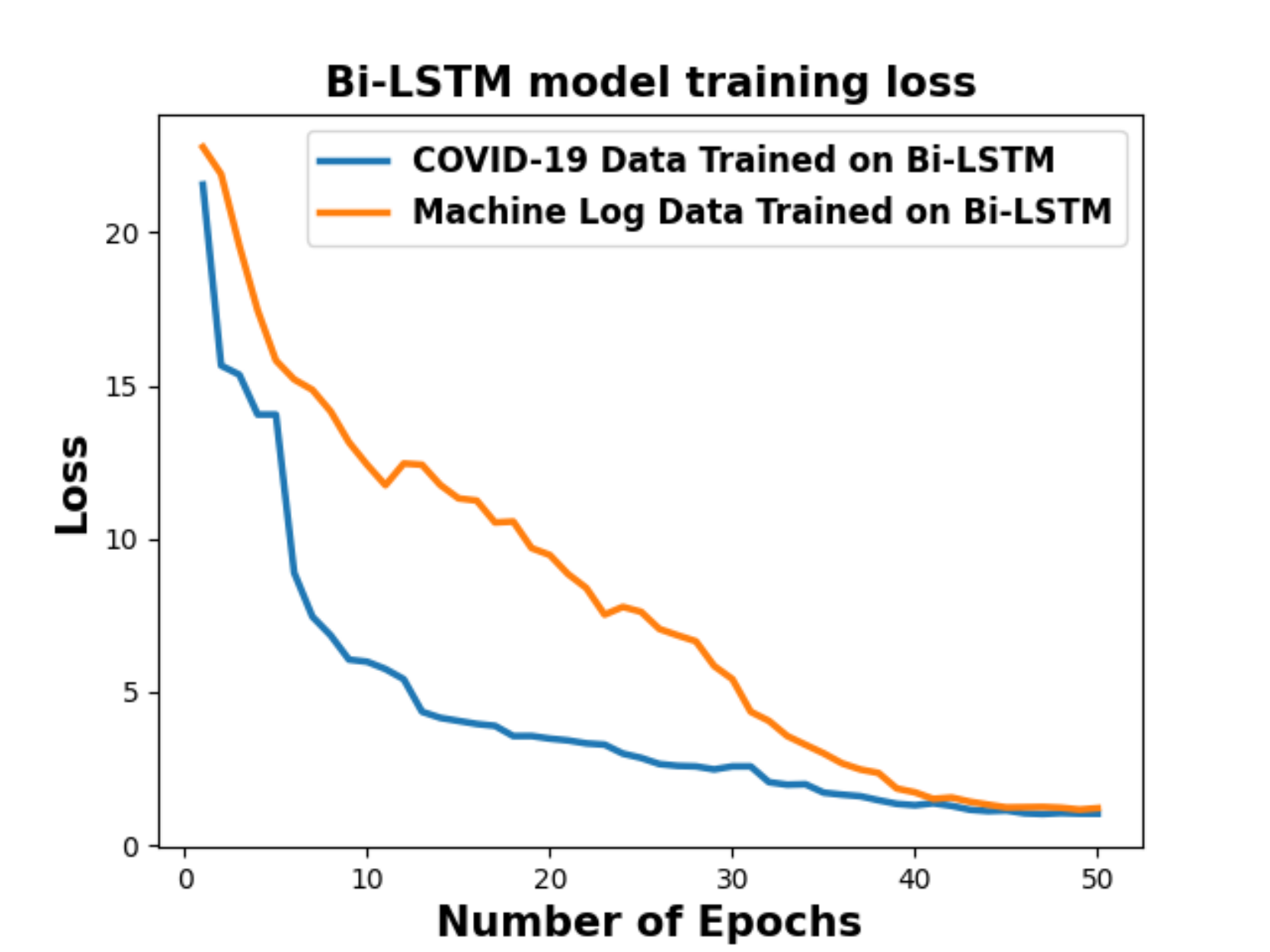}  
}%
\hspace{0pt}
\centering\subfigure[Transformer models]{%
   \includegraphics[width=1.67in]{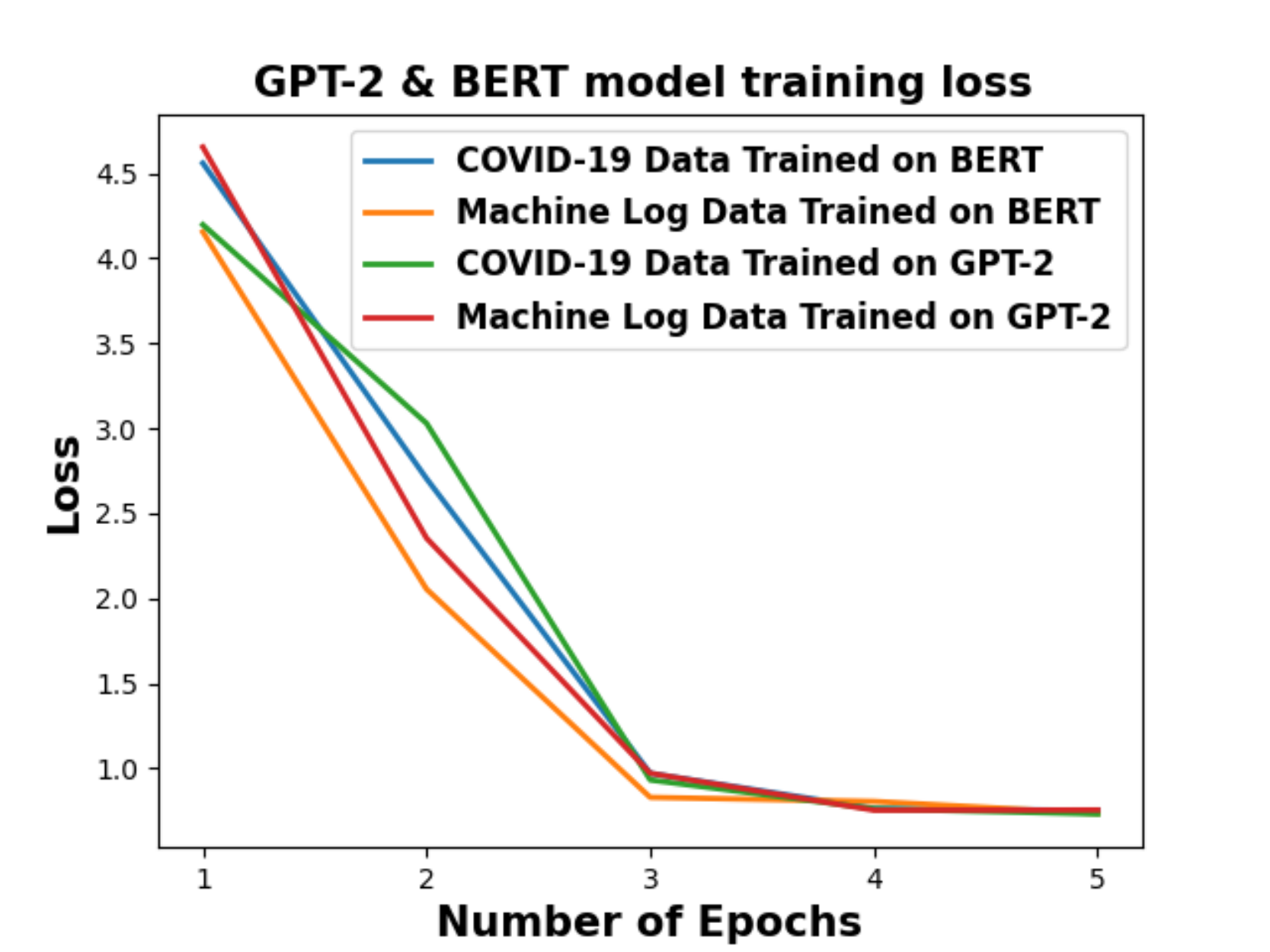}  
}%
\caption{\label{fig:training}
Loss vs. epochs in the training process
}
\end{figure}

\noindent
\textbf{Model Training}
For the training process of both scenarios, Bi-LSTM is relatively slower in converging, requiring around $50$ epochs; while the models leveraging pre-trained GPT-2 small and BERT base are much faster to converge, requiring only around $5$ epochs, as illustrated in Fig.~\ref{fig:training}.

\noindent
 \textbf{Metrics.} We evaluate and compare the accuracy, the storage overhead, and the end-to-end training and inference latency, with all types of schema changes as mentioned in Sec.~\ref{sec:perturbations} applied at the inference stage.  The accuracy of the data integration model is defined as the ratio of the number of super cells that has been predicted with correct target positions and aggregation actions to the total number of super cells in the testing data. 

\noindent
\textbf{Hardware Platform.} For all experiments, if without specification, we use one NVIDIA Tesla V100 GPU from Google Colab. All running times (e.g., training time, inference time) are measured as the average of multiple repeated runs.

\subsection{Coronavirus Data Integration Scenario}
\subsubsection{Experiment Setup} 
We evaluate our system in a COVID-$19$ data integration scenario that is close to the example in Sec.~\ref{sec:intro}. We predict COVID-$19$ trend using daily and regional information regarding the number of vaqarious cases and mobility factors.  Given a set of raw data sources, we need to create a $2$-dimensional target dataset on daily basis. In the target dataset, each row represents coronavirus and mobility information for a state/province on the specific date, and each column represents the state, country, number of confirmed cases, recovery cases, death cases, and the mobility factors regarding workplace, grocery, transit, etc.. The target dataset can be used as inputs to various curve-fitting techniques~\cite{death-prediction-cdc, covid19-models} for COVID-$19$ prediction. 

\noindent
\textbf{Datasets.} We assume the user specifies/recommends a small set of initial data sources. For the first scenario, the user specifies the John Hopkins University's COVID-$19$ github repository~\cite{covid19-jhu} and Google mobility data~\cite{google-mobility}. The statistics about the above source tables are illustrated in Tab.~\ref{tab:table-statistics}. The JHU dataset contains $258$ files with each file representing COVID-$19$ statistics on a specific date. These files have tens of versions, growing from $36$ rows and $6$ attributes to $3956$ rows and $14$ columns.

\noindent
\textbf{Perturbations.} We add perturbations such as random changes to attribute names and values, and replacing attribute names and value tokens by synonyms as described in Sec.~\ref{sec:perturbations} to $58.3\%$ of the attributes in the training data. In addition, we add key expansion changes, which accounts for $18.6\%$ of the rows in the training data. We test the model using JHU-COVID-19 data and Google mobility data collected from Feb 15, 2020 to Oct 6, 2020, as illustrated in Tab.~\ref{tab:table-statistics}.

\begin{table}[h]
\centering
\scriptsize
\caption{\label{tab:table-statistics} Statistics of Relevant Data Sources.}
\begin{tabular}{|r|r|r|r|} \hline
&numFiles&numRows&numCols\\\hline \hline
JHU-COVID19&$258$&$36$-$3956$&$6$-$14$\\ \hline
Google-Mobility&$1$&$2,526,500$&$14$\\ \hline
\end{tabular}
\end{table}

\subsubsection{Overall Results} 
The overall results are illustrated in Tab.~\ref{tab:results}, which show that employing a complex transformer like the pre-trained GPT-2 small and BERT base, we can achieve better accuracy, though more complicated models require significantly more storage space and computational time for training one epoch or inference.

The results also show that with the increase of the granularity of super cells , the required training and testing time will be significantly reduced, while the accuracy will decrease. 

\begin{table}[ht]
\centering
\scriptsize
\caption{\label{tab:results} COVID-19 data integration with different super cell granularity (number of super cells per tuple of target dataset) }
\begin{tabular}{|r|l|r|r|r|r|} \hline
\#supercell&model&accuracy&$T_{train}/epoch$&$T_{predict}$ \\\hline \hline
\multirow{3}{*}{9}&Bi-LSTM&$96.6\%$&$4.8$ sec&$18.9$ sec\\
\cline{2-5}
&GPT-2 small&$99.8\%$&$28.1$sec&$21.7$ sec\\ 
\cline{2-5}
&BERT base&$99.8\%$&$29.3$ sec&$20.6$ sec\\ \hline
\multirow{3}{*}{4}&Bi-LSTM&$92.4\%$&$3.2$ sec&$8.6$ sec\\
\cline{2-5}
&GPT-2 small&$99.6\%$&$9.2$ sec&$6.8$ sec\\ 
\cline{2-5}
&BERT base&$99.6\%$&$9.6$ sec&$6.7$ sec\\ \hline
\multirow{3}{*}{2}&Bi-LSTM&\multicolumn{3}{|c|}{Failed}\\
\cline{2-5}
&GPT-2 small&$99.4\%$&$5.8$ sec&$3.8$ sec\\ 
\cline{2-5}
&BERT base&$99.4\%$&$6.0$ sec&$3.8$ sec\\ \hline

\end{tabular}
\end{table}

\subsubsection{Ablation Study} 
Using the Bi-LSTM model with single-cell representation, we also conducted detailed ablation study as illustrated in Tab.~\ref{tab:ablation}. It shows that handling value format changes (e.g., date format change like 10-06-2020 and 06102020; and different abbreviations of region and subdistricts like AZ and Arizona.) is a main factor for accuracy degradation. Using a customized synonymous dictionary to encode these format changes for adding perturbations to the training data can greatly improve the accuracy compared with using synonyms extracted from Google Knowledge Graph, as illustrated in Tab.~\ref{tab:ablation1}. In addition, we also find that using character-based embedding can significantly outperform word-based embedding, as illustrated in Tab.~\ref{tab:ablation2}.

\begin{table}[h]
\centering
\scriptsize
\caption{\label{tab:ablation} Ablation Study for different schema change types, changes are incrementally added.}
\begin{tabular}{|l|r|} \hline
Testing Cases&Testing Accuracy\\\hline \hline
relevant data with no schema changes&$99.9\%$\\ \hline
irrelevant data&$99.9\%$\\ \hline
changes of two attributes&$99.9\%$\\ \hline
changes of five attributes&$99.9\%$\\ \hline
changes of six attributes and value format changes&$97.5\%$\\ \hline
key expansion&$96.4\%$\\ \hline
\end{tabular}
\end{table}

\begin{table}[h]
\centering
\scriptsize
\caption{\label{tab:ablation1}Accuracy comparison for using synonyms retrieved from Google Knowledge Graph (GKG) and using self-coded synonyms.}
\begin{tabular}{|l|r|r|} \hline
Testing Cases&GKG  &Self-coded  \\\hline \hline
changes of two attributes&$94.5\%$&$99.9\%$\\ \hline
changes of five attributes&$97.5\%$&$99.9\%$\\ \hline
changes of six attributes and region format changes&$82.2\%$&$97.5\%$\\ \hline
\end{tabular}
\end{table}

\begin{table}[h]
\centering
\scriptsize
\caption{\label{tab:ablation2}Accuracy comparison for using different embedding approaches.}
\begin{tabular}{|l|r|} \hline
Testing Cases&accuracy   \\\hline \hline
Word2Vec pretrained on Wikipidia &$71.2\%$\\ \hline
Word2Vec pretrained on Google News &$67.5\%$\\ \hline
fastText pretrained on self-customized corpus  &$99.9\%$\\ \hline
\end{tabular}
\end{table}

\subsubsection{Human Productivity Comparison}
We developed the data integration code using Python and Pandas dataframe to integrate the JHU COVID-19 data collected on Feb 15, 2020 and the time-series Google mobility data. After Feb 15, 2020, the first schema evolution of the JHU COVID-19 data schema that breaks the integration code and causes system downtime, happened on Mar 22, 2020. we invite an experienced software engineer, a Ph.D student, and an undergraduate student to develop the revisions respectively and ask them to deliver the task as soon as possible. We record the time between the task assignment and code submission, as well as the time they dedicated to fixing the issue as they reported. We find that although the reported dedicated time ranges from $15$ to $25$ minutes; the time between the task assignment and code submission ranges from one to three days. This example illustrates the unpredictability of human resources. 
In contrast, our proposed data integration pipeline can smoothly handle schema changes without any interruptions or delays, and requires no human intervention at all.

\noindent
\textbf{Performance of Python-based Integration Code.} We run our Python-based and human-coded data integration pipeline on the aforementioned daily JHU COVID-19 data and Google mobility data in a C4.xlarge AWS instance that has four CPUs and eight gigabytes memory, and it takes $417$ seconds of time on average to integrate data for one day, without considering the time required to fix the pipeline for schema changes. $97\%$ of the time is spent on removing redundant county-level statistics from the relatively large Google mobility file that has $2.5$ millions of tuples. Otherwise the co-existing state-level and county-level statistics in the Google mobility file will confuse the join processing. This observation indicates that with the acceleration of high-end GPU processor, the overall training and inference latency of using a deep learning based pipeline is lower than using the traditional human-centered pipeline. Considering that the training process only needs to be carried out at the beginning and when a concept drift~\cite{tsymbal2004problem} is detected.

\subsection{Machine Log Integration}

\subsubsection{Environment Setup}
Suppose a lab administrator developed a Python tool to integrate various performance metrics of a cluster of MacOS workstations, such as CPU utilization (user, system, idle, wait), memory utilization (cached, buffered, swap), network utilization (input, output), disk utilization (write, read), and so on. The tool collects these metrics by periodically reading the output of an omnipresent shell tool ``top"\footnote{https://linux.die.net/man/1/top} and then perform a union operation for time-series metrics collected from each machine. Now the lab purchased four Ubuntu Linux servers. However, because the ``top" tool's output in Ubuntu is very different from MacOS, the Python tool cannot work with these new Linux machines without additional coding efforts. Such problem is prevalent in machine or sensor data integration, where different devices produced by different manufacturers may use different schemas to describe similar information.

\subsubsection{Overall Results}
The results are illustrated in Tab.~\ref{tab:results1} and Tab.~\ref{tab:results2}, showing that our approach can achieve acceptable accuracy. Particularly, the transformer models can achieve significantly better accuracy than the Bi-LSTM model. For this case, with the increase in super cell granularity (i.e., decrease in number of super cells per target tuple), the accuracy of the Bi-LSTM network is improved, while the accuracy of the transformer-based models is slightly degraded. The transformer-based models can achieve significantly better accuracy, while the computational time required for training (per epoch) and inference is significantly higher. Also the larger of the super cell granularity, the fewer number of training and testing samples. Therefore, the time required for training and testing is also significantly reduced with the increase in the super cell granularity.

\begin{table}[ht]
\centering
\scriptsize
\caption{\label{tab:results1} Machine log data integration with different super cell granularity (number of super cells per tuple of target dataset): Ubuntu and MacOS data used for training, and Android data used for testing}
\begin{tabular}{|r|l|r|r|r|r|} \hline
\#supercell&model&accuracy&$T_{train}/epoch$&$T_{predict}$ \\\hline \hline
\multirow{3}{*}{49}&Bi-LSTM&$82.1\%$&$8.7$ sec&$8.1$ sec\\
\cline{2-5}
&GPT-2 small&$99.6\%$&$52.4$sec&$21.3$ sec\\ 
\cline{2-5}
&BERT base&$99.7\%$&$39.1$ sec&$15.3$ sec\\ \hline
\multirow{3}{*}{25}&Bi-LSTM&$85.3\%$&$5.4$ sec&$3.6$ sec\\
\cline{2-5}
&GPT-2 small&$99.4\%$&$29.3$ sec&$9.8$ sec\\ 
\cline{2-5}
&BERT base&$99.4\%$&$22.7$ sec&$7.6$ sec\\ \hline
\multirow{3}{*}{3}&Bi-LSTM&$91.2\%$&$2.8$ sec&$1.9$ sec\\
\cline{2-5}
&GPT-2 small&$99.0\%$&$12.3$ sec&$4.0$ sec\\ 
\cline{2-5}
&BERT base&$99.1\%$&$13.9$ sec&$3.7$ sec\\ \hline
\end{tabular}
\end{table}

\begin{table}[ht]
\centering
\scriptsize
\caption{\label{tab:results2} Machine log data integration with different super cell granularity (number of super cells per tuple of target dataset): only Ubuntu data used for training, MacOS and Android data used for testing}
\begin{tabular}{|r|l|r|} \hline
\#supercell&model&accuracy \\\hline \hline
\multirow{3}{*}{49}&Bi-LSTM&$63.7\%$\\
\cline{2-3}
&GPT-2 small&$99.5\%$\\ 
\cline{2-3}
&BERT base&$99.6\%$\\ \hline
\multirow{3}{*}{25}&Bi-LSTM&$71.2\%$\\
\cline{2-3}
&GPT-2 small&$99.3\%$\\ 
\cline{2-3}
&BERT base&$99.3\%$\\ \hline
\multirow{3}{*}{3}&Bi-LSTM&$75.6\%$\\
\cline{2-3}
&GPT-2 small&$98.7\%$\\ 
\cline{2-3}
&BERT base&$99.1\%$\\ \hline
\end{tabular}
\end{table}

\subsection{Output Assembling}
\begin{figure}[ht]
\centering\subfigure[100K-latency]{%
   \label{fig:100k-supercell}
   \includegraphics[width=1.62in]{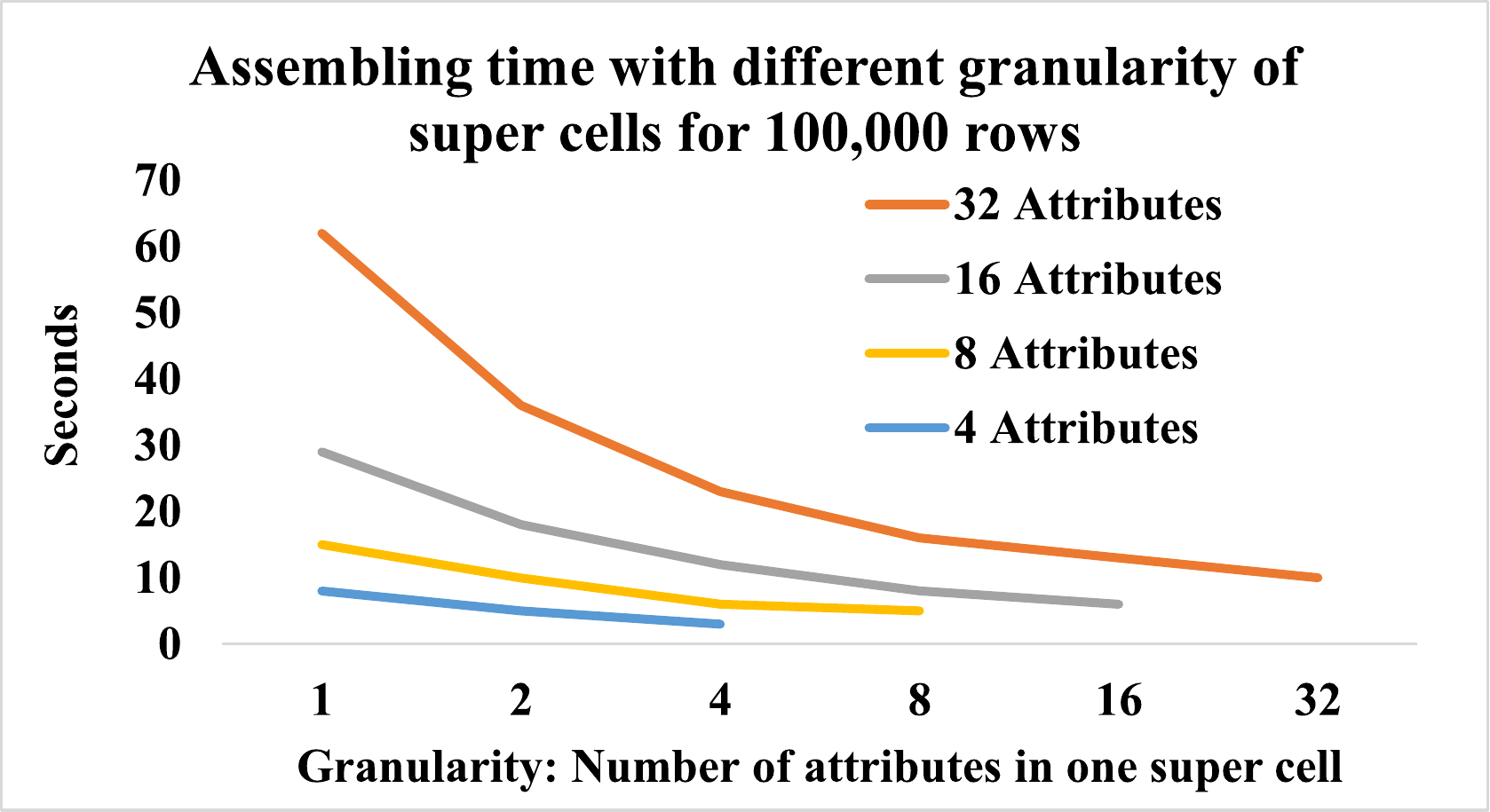}  
}%
\hspace{0.1pt}
\subfigure[100K-breakdown]{%
   \label{fig:100k-breakdown}
   \includegraphics[width=1.62in]{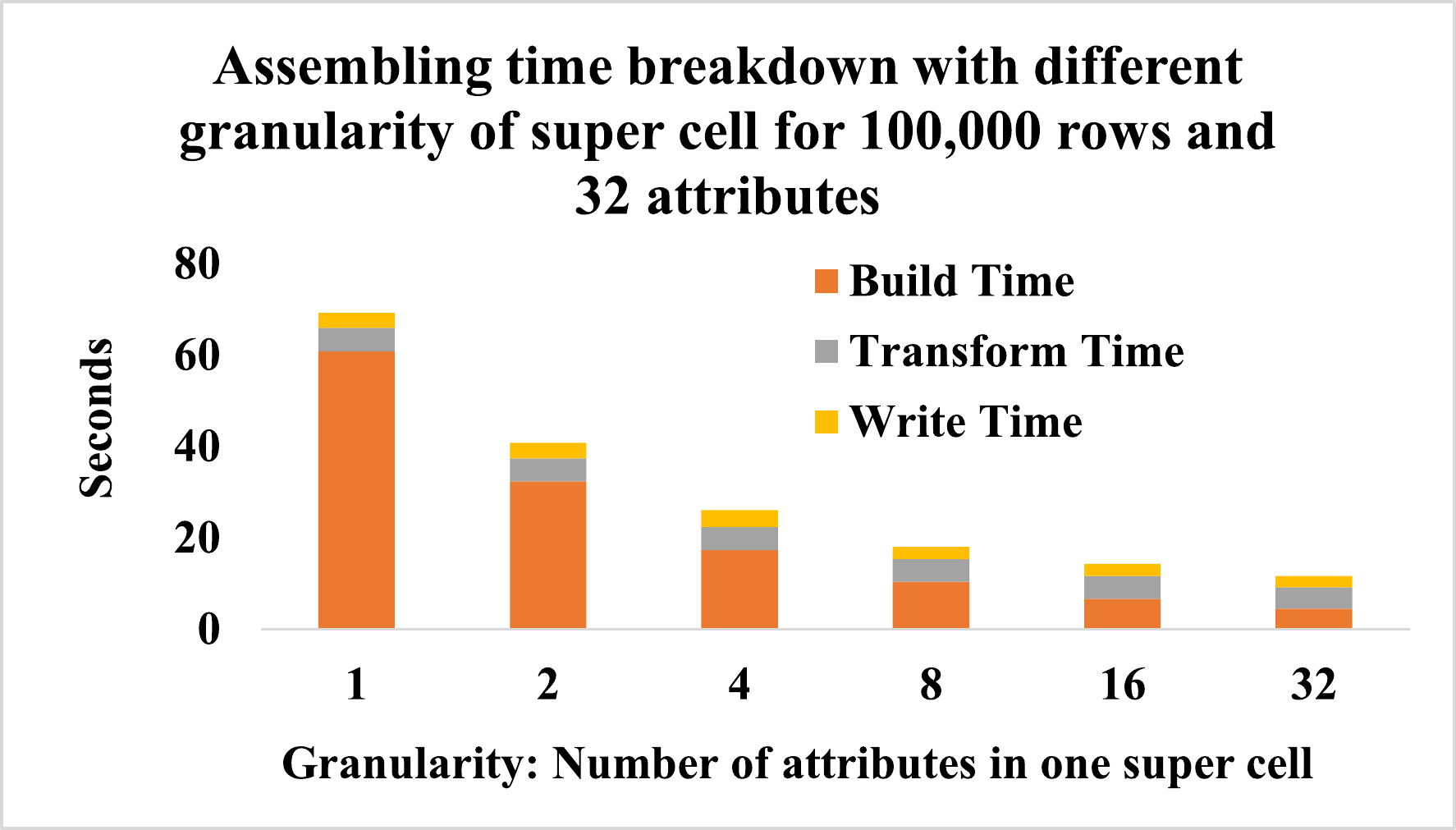}  
}%
\eat{
\subfigure[250K-latency]{%
  \label{fig:250k-supercell}
  \includegraphics[width=1.62in]{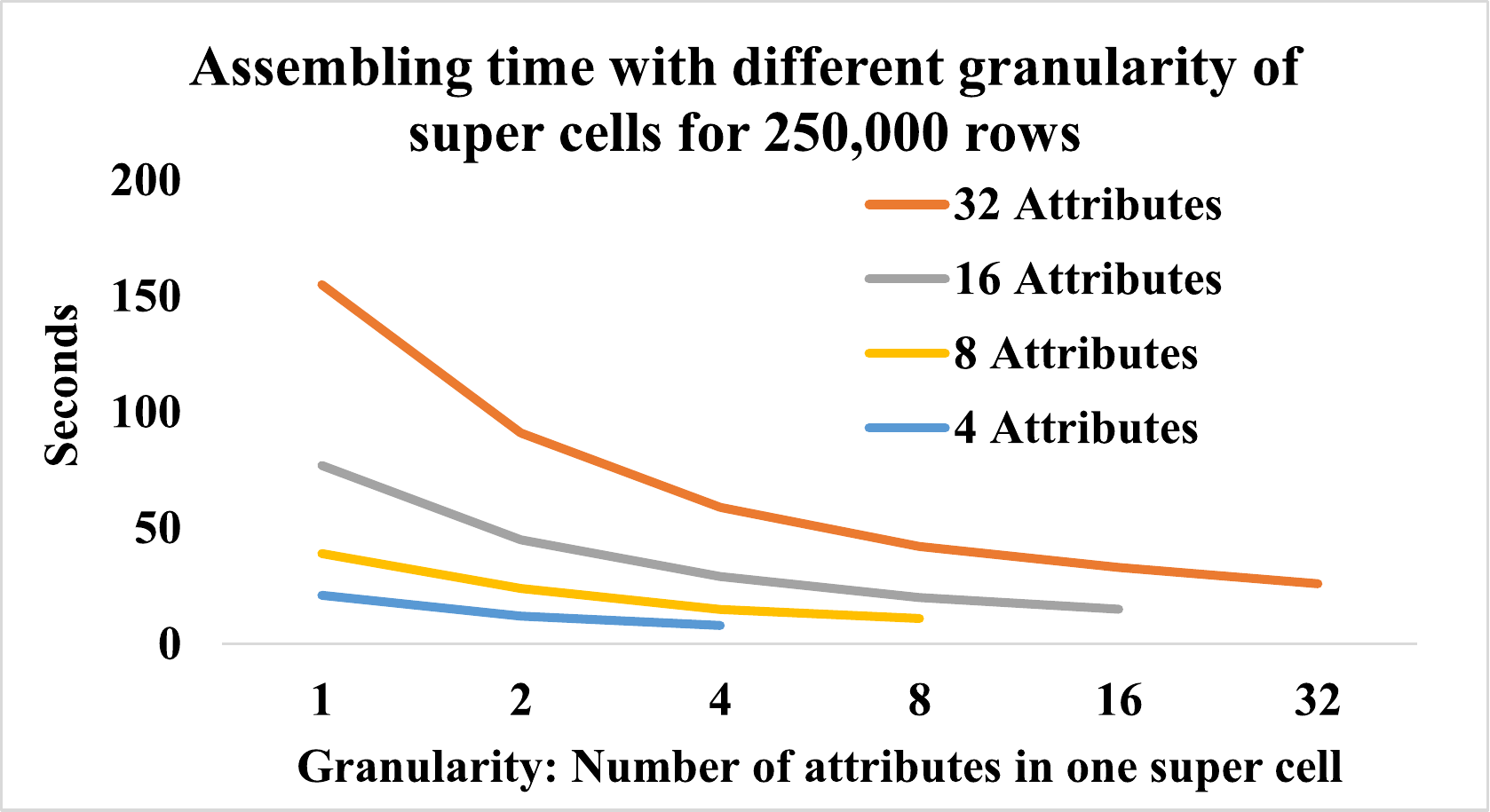}
}%
\hspace{0.1pt}
\subfigure[250K-breakdown]{%
  \label{fig:250k-breakdown}
  \includegraphics[width=1.62in]{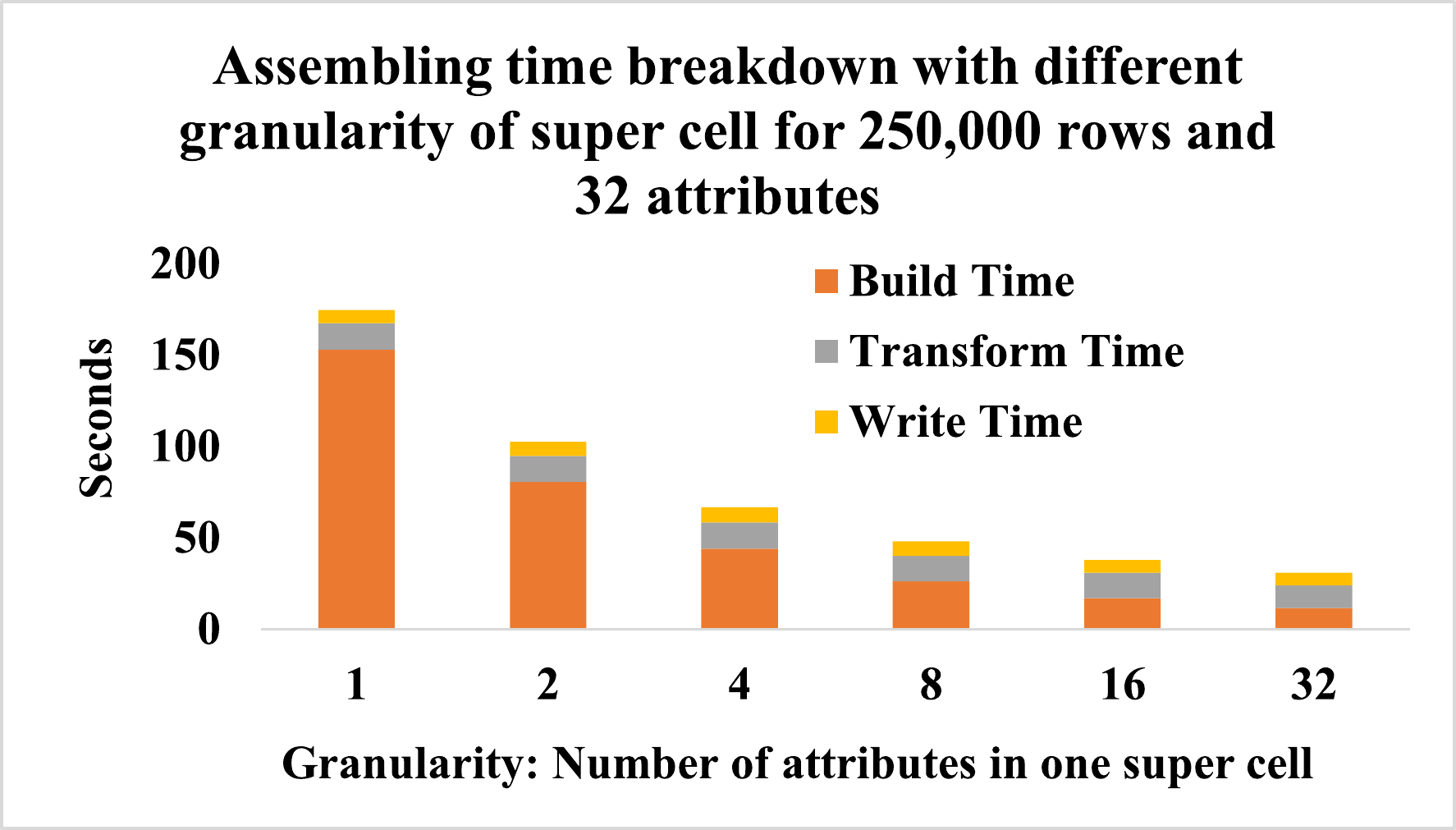}
}%
}
\hspace{0.1pt}
\subfigure[500K-latency]{%
  \label{fig:500k-supercell}
  \includegraphics[width=1.62in]{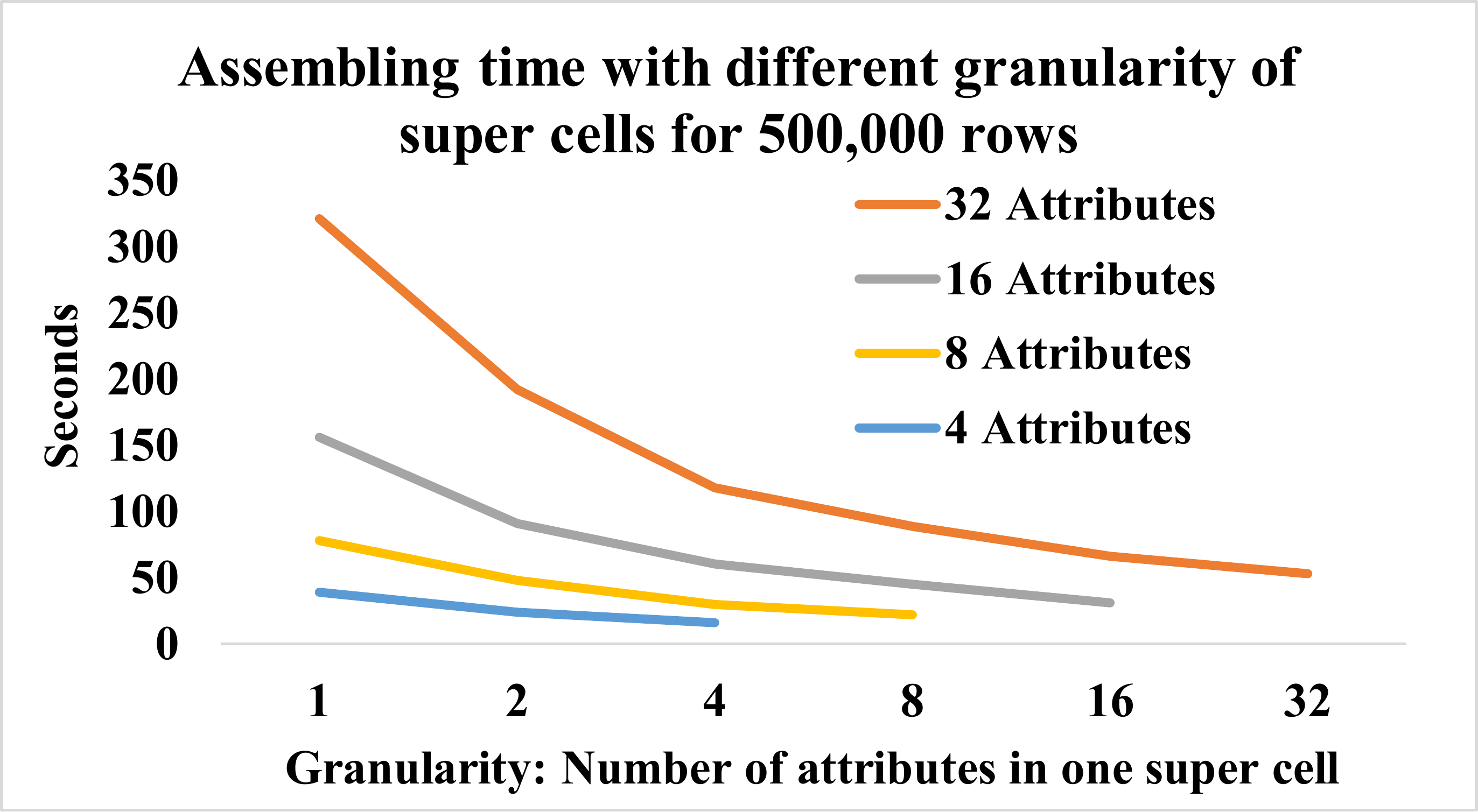}
}%
\hspace{0.1pt}
\subfigure[500K-breakdown]{%
  \label{fig:500k-breakdown}
  \includegraphics[width=1.62in]{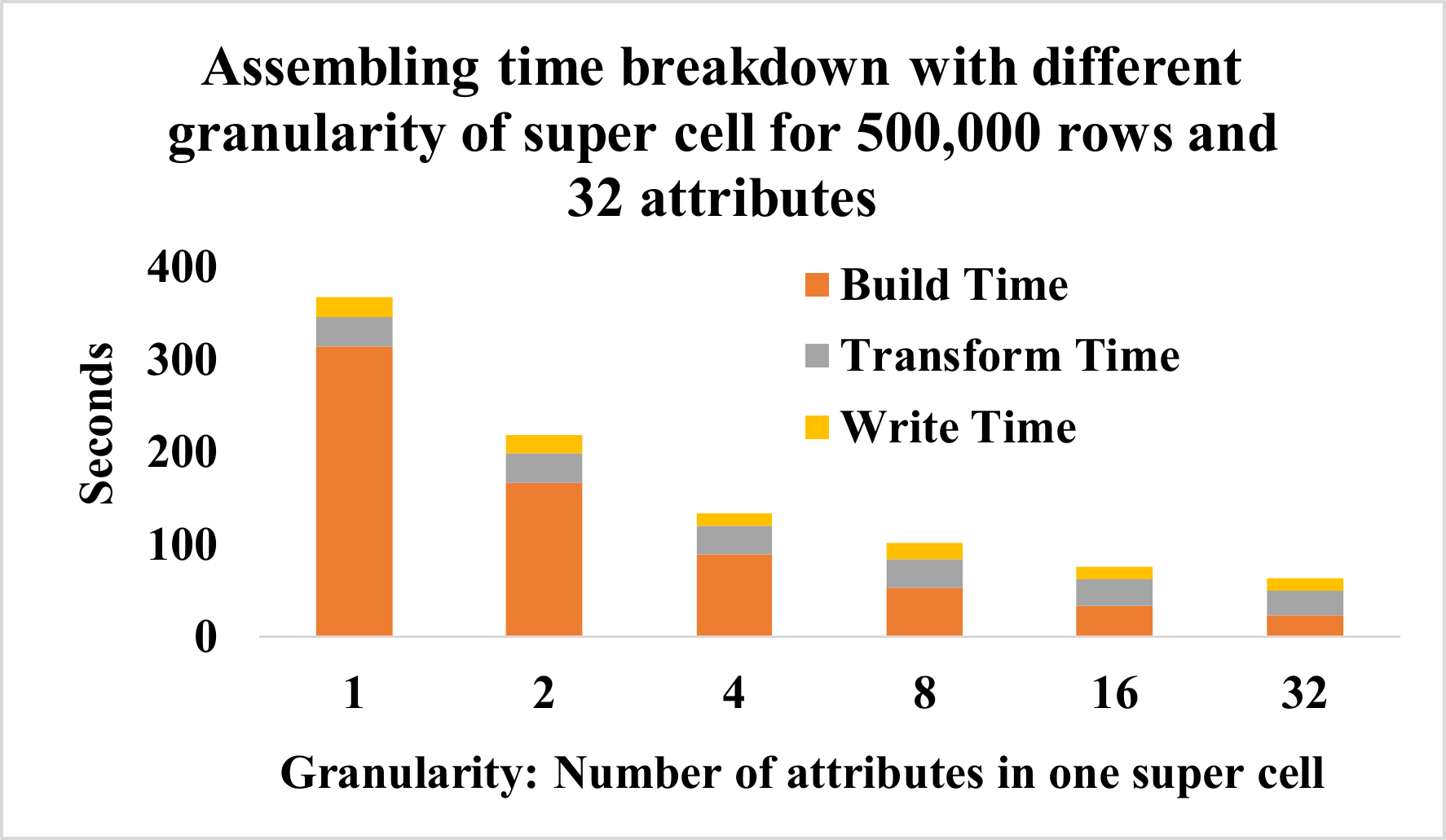}
}%
\eat{
\hspace{0.1pt}
\subfigure[750K-latency]{%
  \label{fig:1million-supercell}
  \includegraphics[width=1.62in]{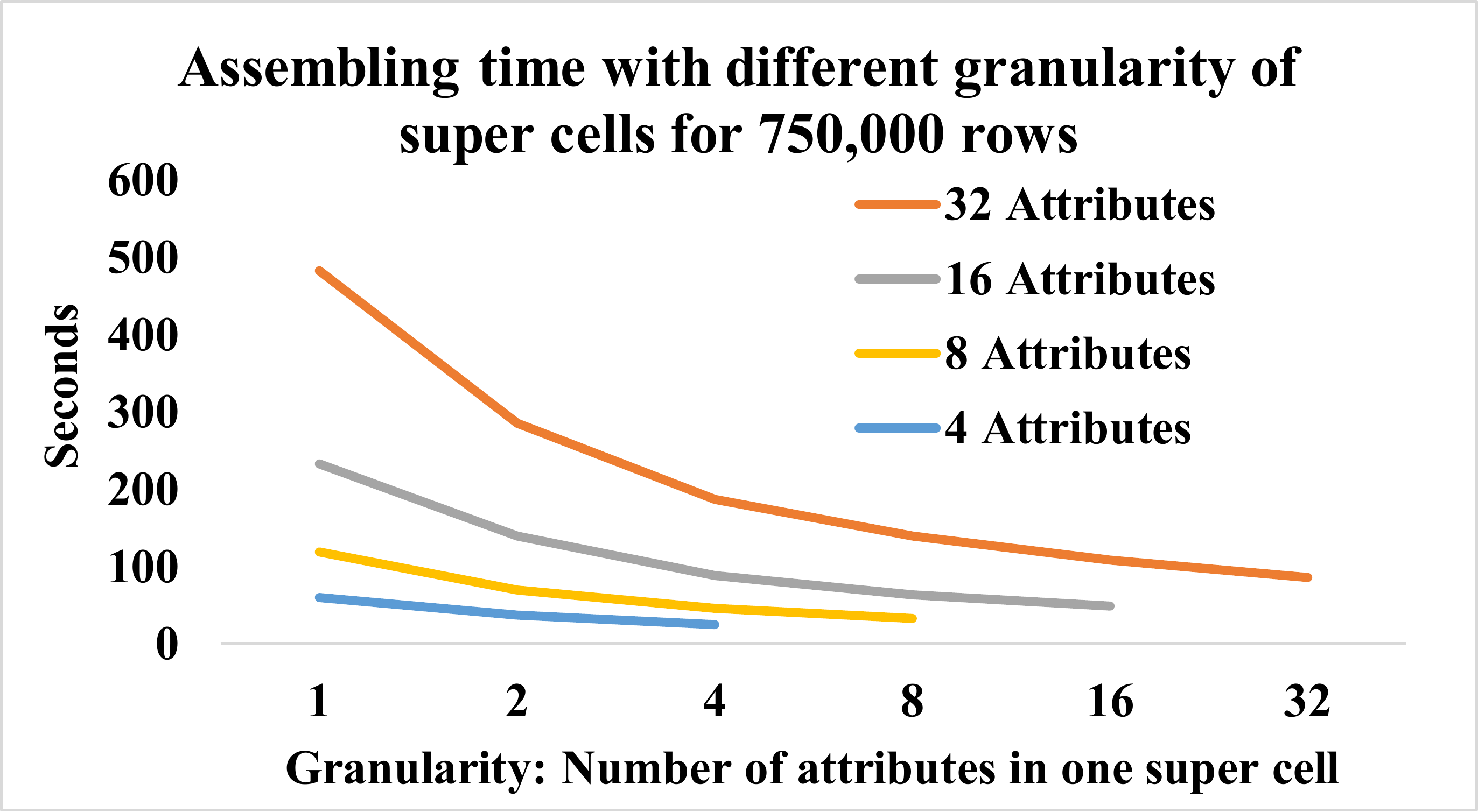}
}%
\hspace{0.1pt}
\subfigure[750K-breakdown]{%
  \label{fig:750k-breakdown}
  \includegraphics[width=1.62in]{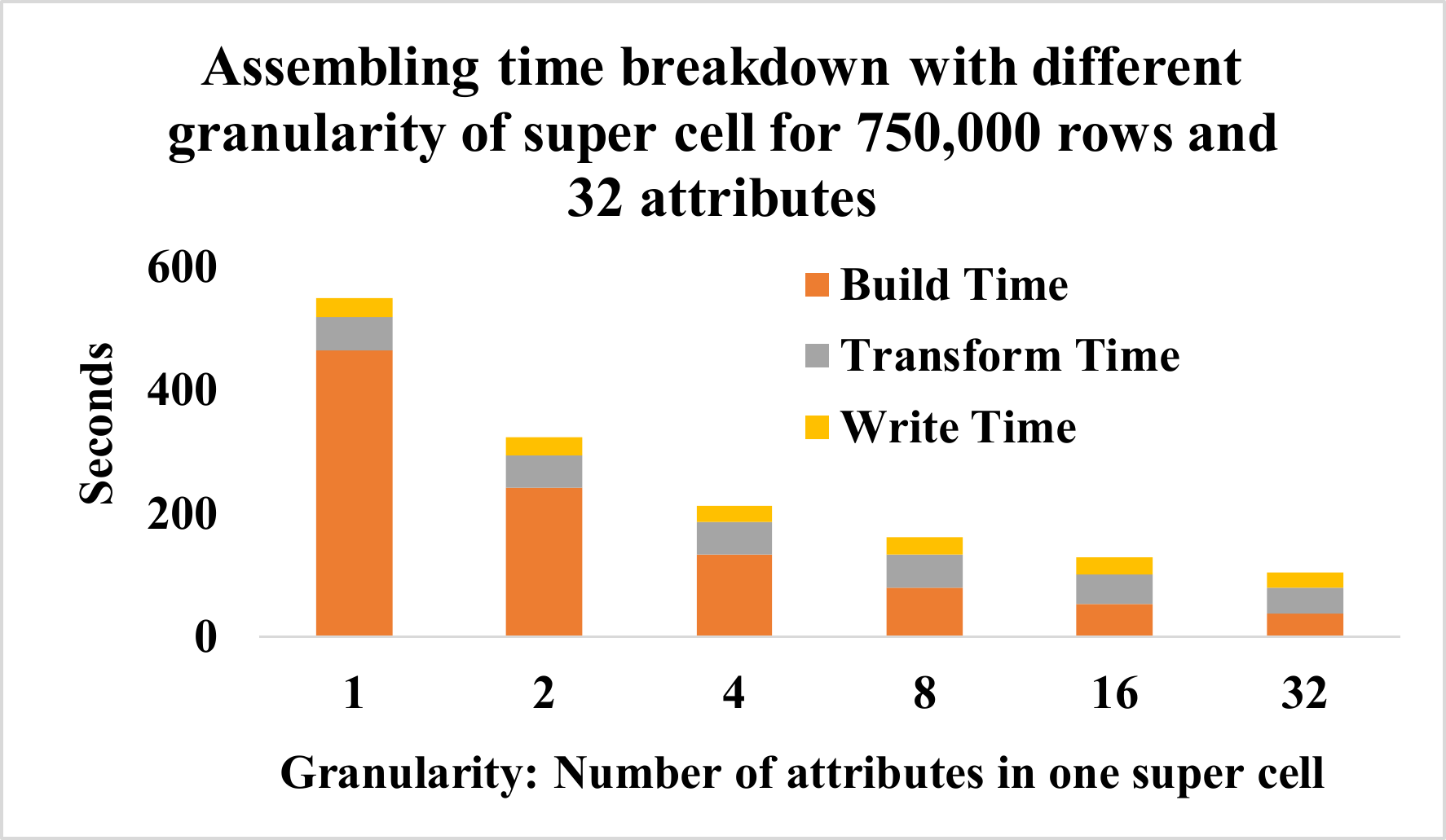}
}%
}
\hspace{0.1pt}
\subfigure[1 million-latency]{%
  \label{fig:1million-supercell}
  \includegraphics[width=1.62in]{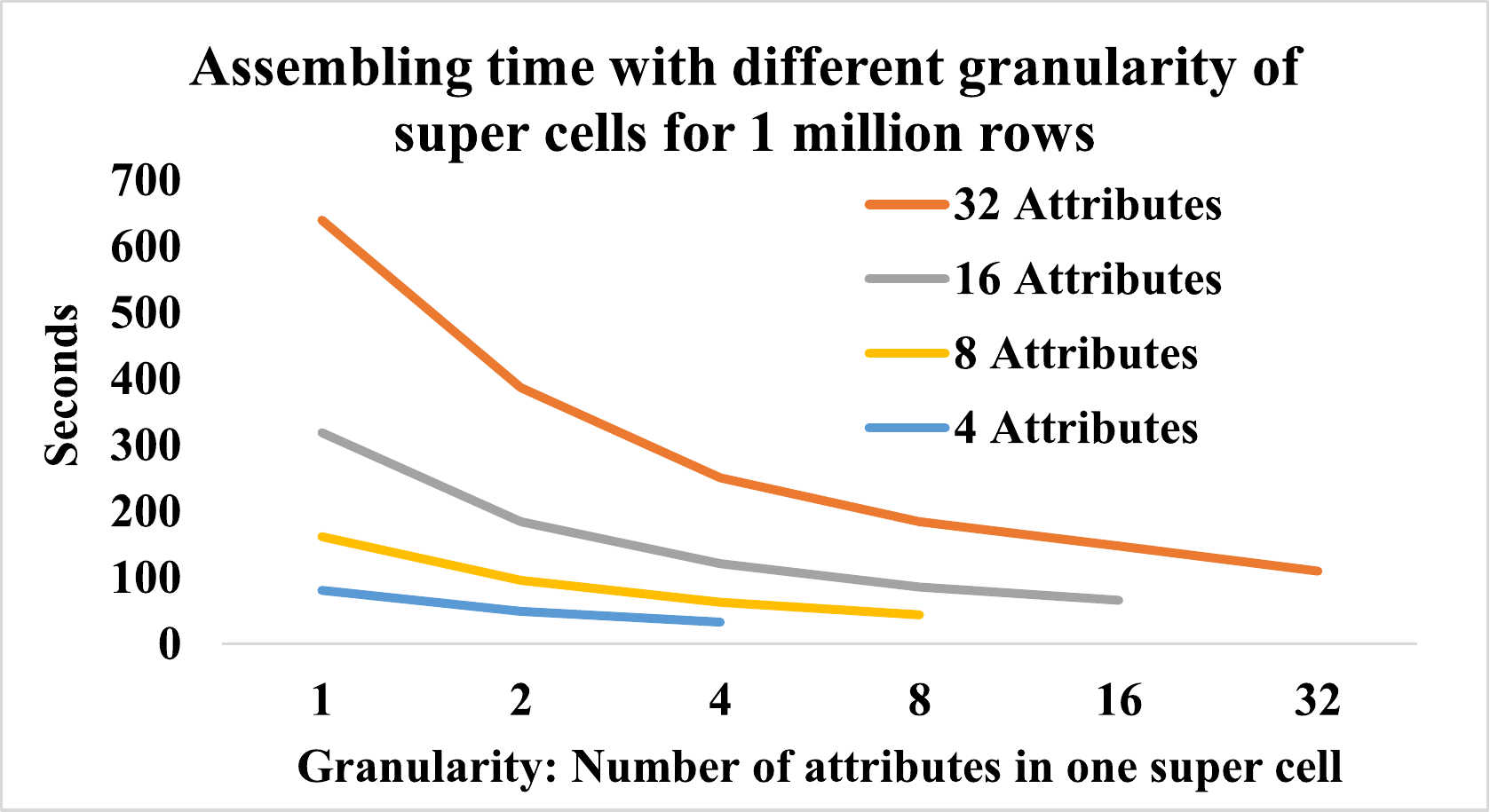}
}%
\hspace{0.1pt}
\subfigure[1 million]{%
  \label{fig:1million-breakdown}
  \includegraphics[width=1.62in]{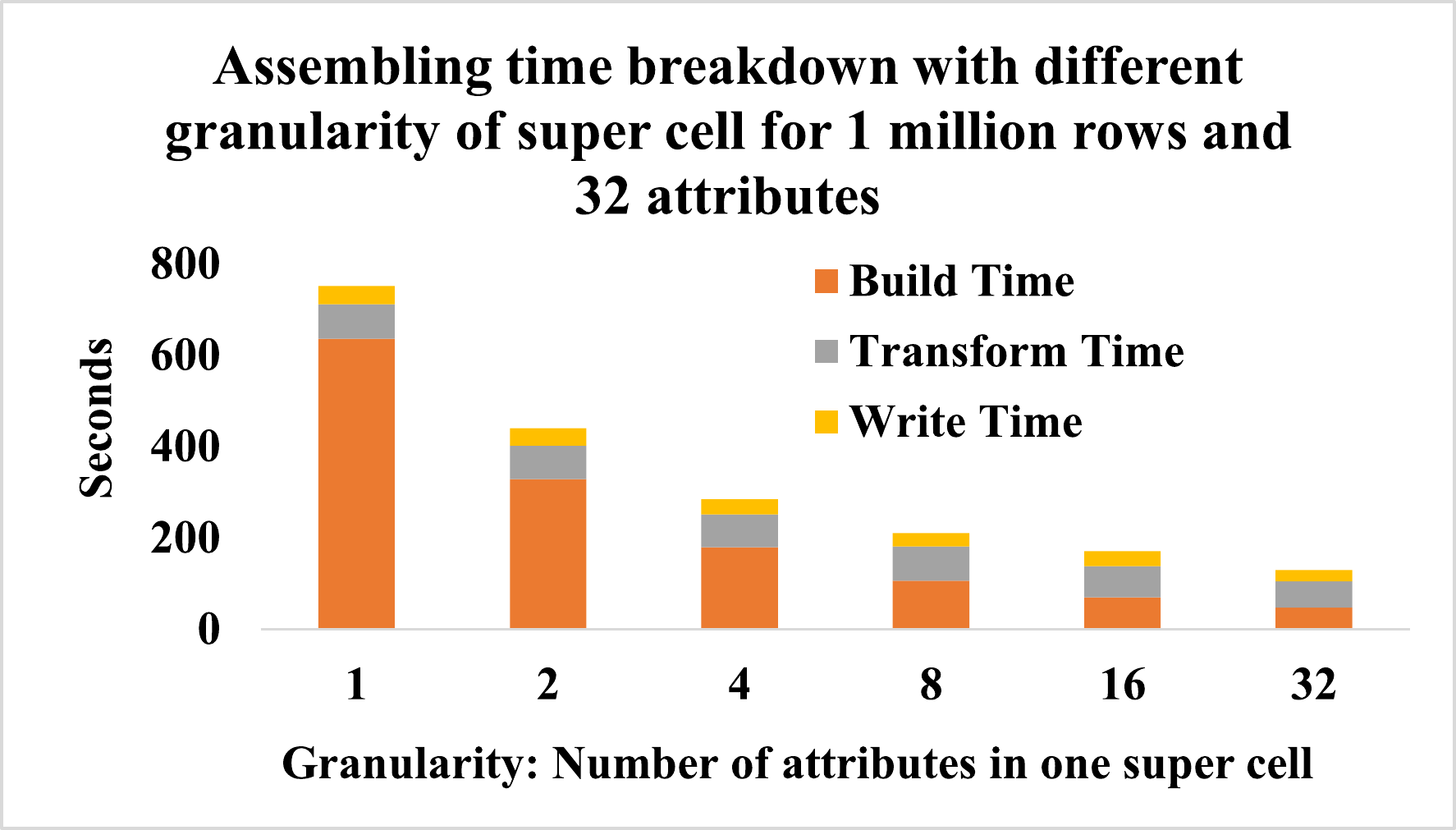}
}

\caption{\label{fig:supercell-assembling}
Influence of different super cell granularity to assembling latency and breakdown analysis (build time: parse and insert cells involved in each prediction to a dictionary. 
transform time: transform the dictionary to a Pandas dataframe. write time: write the Pandas dataframe to a CSV file.). We use different number of attributes and rows for the target table in each experiment.
}
\end{figure}

In this section, we discuss the process of assembling prediction results into tabular files. We mainly measure how the sizes of source datasets, target datasets, and granularity of super cells affect the overall latency of the assembling process. The results are illustrated in Fig.~\ref{fig:supercell-assembling}, which show that increasing super cell granularity will significantly reduce the assembling latency. It indicates that if storage space is not the bottleneck, using a transformer-based model and the largest possible super cell granularity will achieve acceptable accuracy while significantly reducing the computational time required for training, inferences, and assembling.

\section{Related Works}
\label{sec:survey}
\subsection{Handling Schema Evolutions}
Schema evolution in relational database, XML, JSON and ontology has been an active research area for a long time~\cite{rahm2006online, doan2005semantic}. One major approach is through model (schema) management~\cite{bernstein2003applying,bernstein2000data} and to automatically generate executable mapping between the old and evolved schema~\cite{miller2000schema, yu2005semantic, velegrakis2004preserving}. While this approach greatly expands the theoretical foundation of relational schema evolution, it requires application maintenance and may cause undesirable system downtimes~\cite{curino2008graceful}. To address the problem, Prism~\cite{curino2008graceful} is proposed to automate the end-to-end schema modification process by providing DBAs a schema modification language (SMO) and automatically rewriting users' legacy queries. However, Prism requires data migration to the latest schema for each schema evolution, which may not be practical for today's Big Data era. Other techniques include versioning~\cite{klein2001ontology,moon2008managing, sheng2019non}, which avoids the data migration overhead, but incurs version management burden and significantly slows down query performance.  There are also abundant works discussing about the schema evolution problem in NoSQL databases, Polystore or multi-model databases~\cite{storl2020nosql, moller2019query, holubova2019evolution, hillenbrand2019migcast}

\textit{ Most of these works are mainly targeting at enterprise data integration problems and require that each source dataset is managed by a relational or non-relational data store. However the open data sources widely used by today's data science applications are often unmanaged, and thus lack schemas or metadata information~\cite{miller2018open}. A deep learning model, once trained, can handle most schema evolution without any human intervention, and does not require any data migration, or version management overhead. Moreover, today's data science applications are more tolerant to data errors compared to traditional enterprise transaction applications, which makes a deep learning approach promising.}

\eat{
The deep learning approach we propose not only targets at handling most types of schema evolution transparently, but also has a potential to automate the overall process of data discovery, and schema matching in an efficient and economic style. We will now describe prior-arts in these aspects and compare our approach to existing works.}
\subsection{Data Discovery}
Data discovery is to find related tables in a data lake. Aurum~\cite{fernandez2018aurum} is an automatic data discovery system that proposes to build enterprise knowledge graph (EKG) to solve real-world business data integration problems. In EKG, a node represents a set of attributes/columns, and an edge connects two similar nodes. In addition, a hyperedge connects any number of nodes that are hierarchically related. They propose a two-step approach to build EKG using LSH-based and TFIDF-based signatures. They also provide a data discovery query language SRQL so that users can efficiently query the relationships among datasets. Aurum~\cite{fernandez2018aurum} is mainly targeting at enterprise data integration. In recent, numerous works are proposed to address open data discovery problems, including automatically discover table unionability~\cite{nargesian2018table} and joinability~\cite{zhu2016lsh,zhu2019josie}, based on LSH and similarity measures.
Nargesian and et al.~\cite{nargesian2020organizing} propose a Markov approach to optimize the navigation organization as a DAG for a data lake so that the probability of finding a table by any of attributes can be maximized. In the DAG, each node of navigation DAG represents a subset of the attributes in the data lake, and an edge represents a navigation transition. All of these works provide helpful insights from an algorithmatic perspective and system perspective for general data discovery problems. Particularly, Fernandez and et al.~\cite{fernandez2018seeping} proposes a semantic matcher based on word embeddings to discover semantic links in the EKG.

\textit{Our work has a potential to integrate data discovery and schema matching into a deep learning model inference process. We argue that in our targeting scenario, the approach we propose can save significant storage overhead as we only need store data integration models which are significantly smaller than the EKG, and can also achieve better performance for wide and sparse tables. We will prove in the paper that the training data generation and labeling process can be fully automated.}

\subsection{Schema/Entity Matching}
Traditionally, to solve the data integration problem for data science applications, once related datasets are discovered, the programmer will either manually design queries to integrate these datasets, or leverage a schema matching tool to automatically discover queries to perform the data integration. 

There are numerous prior-arts in schema matching~\cite{kimmig2018collective, miller2000schema, gottlob2010schema, ten2013schema}, which mainly match schemas based on metadata (e.g., attribute name) and/or instances. Entity matching (EM)~\cite{christen2012data}, which is to identify data instances that refer to the same real-world entity, is also related. Some EM works also employ a deep learning-based approach~\cite{mudgal2018deep, meduri2020comprehensive, kasai2019low, konda2016magellan,thirumuruganathan2018reuse, zhao2019auto, ebraheem2017deeper}.  Mudgal and et al.~\cite{mudgal2018deep} evaluates and compares the performance of different deep learning models applied to EM with three types of data: structured data, textual data, and dirty data (with missing value, inconsistent attributes and/or miss-placed values). They find that deep learning doesn't outperform existing EM solutions on structured data, but it outperforms them on textual and dirty data.

In addition, to apply schema matching to heterogeneous data sources, it is important to discover schemas from semi-structured or non-structured data. We proposed a schema discovery mechanism for JSON data~\cite{wang2015schema}, among other related works~\cite{discala2016automatic, mior2017nose}.

\textit{Our approach proposes a super cell data model to unify open datasets. We train deep learning models to learn the mappings between the data items in source datasets and their positions as well as aggregation modes in the target table. If we see the context of a super cell in the source as an entity, and the target position of the super cell as another entity, the problem we study in this work shares some similarity with the entity matching problem. The distinction is that the equivalence of two "entities" in our problem is determined by users' data integration logic, while general entity matching problem does not have such constraints.} 

\subsection{Other Related Works} Thirumuruganathan and et al.~\cite{thirumuruganathan2018data} discuss various representations for learning tasks in relational data curation. Cappuzzo and et al.~\cite{cappuzzo2020creating} further propose an algorithm for obtaining local embeddings using a tripartite-graph-based representation for data integration tasks such as schema matching, and entity matching on relational database. We are mainly targeting at open data in CSV, JSON and text format and choose to use a super cell based representation. These works can be leveraged to improve the super cell representation and corresponding embeddings proposed in this work.

\section{Conclusion}
In this work, we propose an end-to-end approach based on deep learning for periodical extraction of user expected tables from fast evolving data sources of open datasets. We further propose a relatively stable super cell based representation to embody the fast-evolving source data and to train models that are robust to schema changes by automatically injecting schema changes (e.g., dimension pivoting, attribute name changes, attribute addition/removal, key expansion/contraction, etc.) to the training data. We formalize the problem and conduct experiments on integration of open COVID-$19$ data and machine log data. The results show that our proposed approach can achieve acceptable accuracy. In addition, by applying our proposed approach, the system will not be easily interrupted by schema changes and no human intervention is required for handling most of the schema changes.

\bibliographystyle{abbrv}
\bibliography{refs}  

\end{document}